\documentclass[%
twocolumn,
 amsmath,amssymb,
 aps,authoryear
]{aastex631}
\usepackage{graphicx}
\usepackage{ulem}
\usepackage{epstopdf}
\usepackage{amsmath}
\usepackage{amssymb}
\usepackage{mathtools}
\usepackage{mathrsfs}
\usepackage{bm}
\setcitestyle{authoryear,round} 
\usepackage{color}
\usepackage{xcolor}
\usepackage{textcomp}
\usepackage{gensymb}
\usepackage{hyperref}
\shorttitle{Large-scale Structure and Turbulence in Inner Solar Wind: Global Model vs. PSP Observations}
\shortauthors{Chhiber et al.}
\newcommand{\rs}{\text{R}_\odot}

\newcommand{\bb}{\bm{B}}
\newcommand{\vb}{\bm{v}}

\newcommand{\psp}{\textit{PSP}}
\newcommand{\Rb}{\bm{\mathcal{R}}}
\newcommand{\ddt}[1]{\frac{\partial #1}{\partial t}}
\newcommand{\varepsilonb}{\mbox{\boldmath$\varepsilon$}}
\newcommand{\Omegab}{\mbox{\boldmath$\Omega$}}
\begin{document}
\title{Large-scale Structure and Turbulence Transport in the Inner Solar Wind -- Comparison of Parker Solar Probe's First Five Orbits with a Global 3D Reynolds-averaged MHD Model}

\author[0000-0002-7174-6948]{Rohit Chhiber}
\affiliation{Department of Physics and Astronomy, Bartol Research Institute, University of Delaware, Newark, DE 19716, USA}
\affiliation{NASA Goddard Space Flight Center, Greenbelt, MD 20771, USA}
\correspondingauthor{Rohit Chhiber}
\email{rohit.chhiber@nasa.gov}
\email{rohitc@udel.edu}

\author[0000-0002-0209-152X]{Arcadi~V. Usmanov}
\affiliation{Department of Physics and Astronomy, Bartol Research Institute, University of Delaware, Newark, DE 19716, USA}
\affiliation{NASA Goddard Space Flight Center, Greenbelt, MD 20771, USA}

\author[0000-0001-7224-6024]{William~H. Matthaeus}
\affiliation{Department of Physics and Astronomy, Bartol Research Institute, University of Delaware, Newark, DE 19716, USA}

\author[0000-0002-5317-988X]{Melvyn~L. Goldstein}
\affiliation{NASA Goddard Space Flight Center, Greenbelt, MD 20771, USA}
\affiliation{University of Maryland Baltimore County, Baltimore, MD 21250, USA}

%
\begin{abstract}
Simulation results from a global magnetohydrodynamic model of the solar corona and solar wind are compared with Parker Solar Probe (PSP) observations during its first five orbits. The fully three-dimensional model is based on Reynolds-averaged mean-flow equations coupled with turbulence transport equations. The model includes the effects of electron heat conduction, Coulomb collisions, turbulent Reynolds stresses, and heating of protons and electrons via a turbulent cascade. Turbulence transport equations for average turbulence energy, cross helicity, and correlation length are solved concurrently with the mean-flow equations. Boundary conditions at the coronal base are specified using solar synoptic magnetograms. Plasma, magnetic field, and turbulence parameters are calculated along the PSP trajectory. Data from the first five orbits are aggregated to obtain trends as a function of heliocentric distance. Comparison of simulation results with PSP data shows good agreement, especially for mean-flow parameters. Synthetic distributions of magnetic fluctuations are generated, constrained by the local rms turbulence amplitude given by the model. Properties of this computed turbulence are compared with PSP observations.
\end{abstract}


\section{Introduction} \label{sec:intro}
The Parker Solar Probe 
\citep[PSP;][]{fox2016SSR}
has recently completed its eighth solar encounter and the rate and significance of discoveries made thus far  \citep[e.g.,][]{bale2019Nat,kasper2019Nat,mccomas2019Nat,howard2019Nat} are expected to increase as the spacecraft orbit moves further inward. Novel observations of magnetic switchbacks,
plasma jets, dust, 
and energetic particle populations have energized heliophysics research. 
The accumulation
of these data also afford an opportunity to 
examine trends and average properties 
based on the observations
themselves, assisted by comparison with 
large-scale three dimensional (3D) modeling. 
Here we carry out such comparisons, based on the first five complete PSP orbits spanning heliocentric distances from 28 to 200 \(\rs\).
The framework that we employ is a full 3D
magnetohydrodynamic (MHD) model that follows both resolved (typically large-scale), plasma flows and electromagnetic fields,
as well as local statistical properties of unresolved MHD-scale turbulence.

The paper is organized as follows: 
Section \ref{sec:theory}
provides an overview of the 
numerical model, 
emphasizing the physical content of the 
Reynolds-averaged equations
that are coupled to 
a turbulence transport model. The subsections provide details of the model parameters employed here with boundary conditions based on magnetograms.   
Section \ref{sec:data} describes the data analysis procedures employed, and in Section \ref{sec:tser} direct comparisons of the model results with PSP observational datasets are presented for the first five orbits. A different view of the model comparisons with PSP data is provided in Section \ref{sec:radial} where data from the first five orbits are aggregated to extract radial trends that are then compared to five model runs with  corresponding  magnetogram boundary conditions. 
Section \ref{synthetic}
explores the possibility that fluctuations 
and variability observed in particular orbits can be explained using 
synthetic data with properties constrained by the turbulence modeling solutions. 
Section \ref{sec:discuss}
summarizes the comparisons and what has been learned 
from those comparisons. 
Particular points 
of emphasis
are the physics that is included in the 
model, its successes and deficiencies, 
and 
possible directions for improvement. \added{Appendix \ref{sec:app} uses PSP observations to briefly investigate the structural similarity of autocorrelations of different turbulent fields, an assumption that is integral to the turbulence transport model used here.}

\section{Model and Underlying Theory}\label{sec:theory}

The modeling approach we employ is based
on the premise that the large-scale
features of the solar wind are 
well described by two-fluid magnetohydrodynamics (MHD), 
 noting that internal energy 
is meaningfully separated into electron and proton fluid ingredients \citep{cranmer2009ApJ}. 
The dynamics of these large-scale features is determined to a significant degree, but not completely, by 
boundary conditions, so that 
information flow is mainly along characteristics.  Features separated in angle by more than a few tens of degrees do not communicate well \citep[see, e.g.,][]{matthaeus1986prl}. 
At smaller scales the system is subject to local turbulence interactions, which, although formally deterministic, are conveniently approximated by a statistical treatment, such as Reynolds averaging  \citep[e.g.,][]{mccomb1990physics,usmanov2018}.

In particular, 
the model includes a global 3D compressible MHD two-fluid (protons and electrons) treatment of the solar corona and solar wind  together with turbulent transport and heating from the coronal base to interstellar space \citep{usmanov2016four}.
Mean-field (Reynolds-averaged) solar wind equations are solved simultaneously with model equations that describe turbulence transport, which is represented by equations for average turbulence energy, normalized cross-helicity, and the correlation length of magnetic fluctuations.

\subsection{Reynolds Averaging} \label{sec:rey}

We employ a Reynolds-Averaged Navier-Stokes (RANS) approach, based on Reynolds decomposition  \citep{mccomb1990physics}. 
Physical fields, e.g., $\tilde{\mathbf{a}}$, are separated into a mean and a fluctuating component:
\begin{equation}
\tilde{\mathbf{a}} = \mathbf{a}+\mathbf{a'}, \label{eqn:RA}
\end{equation}
making use of an averaging operation: $\mathbf{a} = \langle \tilde{\mathbf{a}} \rangle$. 
The ensemble average is associated with the resolved 
large scales of motion. 
Then $\mathbf{a'}$ is a fluctuating component, \textit{here assumed to be of arbitrary amplitude}, random, and residing at small scales. By construction, $\langle \mathbf{a'} \rangle = 0$. Application of this decomposition to the MHD equations, together with a set of approximations appropriate to observed solar wind, leads to a set of mean-flow equations that are coupled to small-scale fluctuations via appropriate closures, along with an \textit{additional set of equations} that describe the statistics of turbulence as discussed below. 
For more background, see
\cite{usmanov2011solar,usmanov2012three,usmanov2014three,usmanov2016four,chhiber2017ApJS230,usmanov2018}, and \cite{chhiber2019psp2}.

\subsection{Two-Fluid Reynolds-Averaged MHD Equations 
with Turbulence Transport} \label{sec:2fluid}

We describe the electrons and protons 
by fluid equations with separate energy equations. Furthermore, we assume that their bulk velocity is the same \citep{isenberg1986JGR}. The Reynolds decomposition [Equation \eqref{eqn:RA}] applied to the time-dependent two-fluid MHD equations in the frame of reference corotating with the Sun yields the following set of equations for the large-scale mean flow \citep{usmanov2018}:
\begin{equation}
\ddt{N_S} + \nabla\cdot(N_S\vb) = 0, \label{nseq1}
\end{equation}
\begin{eqnarray}
\lefteqn{\hspace{-1.5cm}\ddt{(\rho\vb)}\, + \,\nabla\cdot\left[\rho\vb\vb
  - \frac{1}{4\pi}\bb\bb + \left(P_S + P_E
  + \underbracket[.5pt]{\frac{\langle B'^2\rangle}{8\pi}} + \frac{B^2}{8\pi}\right){\bf I}
  + \underbracket[.5pt]{\Rb}\right]} \nonumber \\
& & \hspace{-1cm} + \,\rho\left[\frac{GM_{\sun}}{r^2}{\hat {\bf r}}
  + 2\Omegab\times\vb + \Omegab\times(\Omegab\times{\bf r})\right]
  = 0, \label{moeq2}
\end{eqnarray}
\begin{equation}
\ddt{\bb} = \nabla\times(\vb\times{\bf B} + \underbracket[.5pt]{\sqrt{4\pi\rho}\varepsilonb_m}),
  \label{ineq2}
\end{equation}
\begin{equation}
\ddt{P_S} + (\vb\cdot\nabla)P_S + \gamma P_S\nabla\cdot\vb
  = (\gamma - 1)\left(\frac{P_E - P_S}{\tau_{SE}} + \underbracket[.5pt]{f_p Q_T}\right), \label{pseq1}
\end{equation}
\begin{align}
  \ddt{P_E} +& (\vb\cdot\nabla)P_E + \gamma P_E\nabla\cdot\vb 
  = \nonumber \\
  & (\gamma - 1)\left[\frac{P_S - P_E}{\tau_{SE}} - \nabla\cdot{\bf q_E}
  + \underbracket[.5pt]{(1 - f_p)Q_T} \right], \label{peeq1}    
\end{align}
where underbracketed terms represent the influence of turbulence on the mean flow.
The independent variables are heliocentric position vector $\bf r$ and time $t$. 
Dependent variables are velocity in the corotating frame $\vb$, magnetic field $\bb$, number density $N_S$ and the thermal pressure $P_S$ of solar wind (thermal) protons, and thermal pressure of electrons $P_E$. All pressures are assumed to be isotropic. We neglect the electron mass $m_e$ compared with the proton mass $m_p$. Consequently, the mass density is $\rho = m_p N_S$. Parameters appearing in the equations are the sidereal solar rotation rate $\Omegab$, the gravitational constant $G$, adiabatic index $\gamma$ (= 5/3), solar mass $M_{\sun}$, a time scale of Coulomb collisions $\tau_{SE}$ between protons and electrons, and fraction of turbulent energy absorbed by protons $f_p$. The unit vector in the radial direction is $\hat{\bm{r}}$, and {\bf I} is the unit matrix. The electron heat flux ${\bf q}_E$ is taken to be collisional \citep{spitzer1953PhRv} below $5-10~R_\odot$ \citep{chhiber2016solar}, and Hollweg's collisionless model \citep{hollweg1974JGR79,hollweg1976JGR} is used above those heights. \added{In the collisionless model ${\bf q}_E$ is directed along the magnetic field, which implies that the electron heat flux is dominated by the electron strahl \citep[e.g.,][]{Verscharen2019LRSP}}. $Q_T$, defined below, is the source of energy deposition/extraction due to turbulent dissipation. The time scale of Coulomb collisions between protons and electrons can be written as $\tau_{SE} = 1/\nu_E$, where \(\nu_E = [8(2\pi m_e)^{1/2}e^4 N_E\ln\Lambda]/[3m_p(k_B T_E)^{3/2}]
\) is the electron-proton collision rate \citep{hartle1968ApJ151} and \( \ln\Lambda = \ln\left[\frac{3(k_B T_E)^{3/2}}{2\pi^{1/2}e^3 N_E^{1/2}}\right]\) is the Coulomb logarithm. Here $e$ is the elementary charge, \(N_E = N_S\) is number density of electrons, and $k_B$ is the Boltzmann constant.

The mean field momentum equation \eqref{moeq2} and induction equation \eqref{ineq2} are coupled to small-scale fluctuations through the Reynolds stress tensor $\Rb = \rho\langle\vb'\vb' - {\bf b}'{\bf b}'\rangle$, and the turbulent electric field $\varepsilonb_m = \langle\vb'\times{\bf
b}'\rangle$, where ${\bf v}'$ and ${\bf b}' =
\bb'/\sqrt{4\pi\rho}$ are the fluctuations in velocity and magnetic fields, respectively. Note that we have neglected density and pressure fluctuations \citep{matthaeus1990JGR}.  In the present implementation we also neglect the turbulent electric field \(\varepsilon_m\) and off-diagonal terms in the Reynolds stress tensor \citep[see][]{usmanov2018}; all other turbulence terms appearing in the dynamical equations are retained.
\subsection{Turbulence Transport Equations} \label{sec:trans}
Transport equations for fluctuations in the rotating frame are obtained by subtracting the mean field equations from the full MHD equations and averaging 
variances and correlations of the differences \citep{matthaeus1994JGR,usmanov2014three}:

\begin{align}
\ddt{Z^2} +& (\vb\cdot\nabla)Z^2 + \frac{Z^2(1 - \sigma_D)}{2}\nabla\cdot\bm{u}
  + \frac{2}{\rho}\Rb\colon\nabla\bm{u} \nonumber \\
  +& 2\varepsilonb_m\cdot(\nabla\times{\bf V}_A) - ({\bf V}_A\cdot\nabla)(Z^2\sigma_c) \nonumber \\
  +& Z^2\sigma_c\nabla\cdot{\bf V}_A = - \frac{\alpha f^{+}(\sigma_c)Z^3}{\lambda}, \label{z2eq}
\end{align}
\begin{align}
\ddt{(Z^2\sigma_c)} +& (\vb\cdot\nabla)(Z^2\sigma_c)
  - ({\bf V}_A\cdot\nabla)Z^2 + \frac{Z^2\sigma_c}{2}\nabla\cdot\bm{u} \nonumber \\
  +& \frac{2}{\rho}\Rb\colon\nabla{\bf V}_A
+ 2\varepsilonb_m\cdot(\nabla\times\bm{u}) \nonumber \\
  +& (1 - \sigma_D)Z^2\nabla\cdot{\bf V}_A
  = - \frac{\alpha f^{-}(\sigma_c)Z^3}{\lambda}, \label{z2seq}
\end{align}
\begin{equation}
\ddt{\lambda} + ({\bf v}\cdot\nabla)\lambda = \beta f^{+}(\sigma_c)Z,
  \label{lameq}
\end{equation}
where $Z^2 = \langle v'^2 + b'^2\rangle$ is twice the turbulence energy per unit mass, $\lambda$ is the correlation length of turbulent fluctuations, $\sigma_c = 2\langle\vb'\cdot{\bf b}'\rangle Z^{-2}$ is the normalized cross helicity, and $\sigma_D = \langle v'^2 - b'^2\rangle Z^{-2}$ is the (constant) normalized energy difference. Other notations are:
velocity in the inertial frame $\bm{u} = \vb + \Omegab\times{\bf r}$, Alfv\'en velocity ${\bf V}_A = \bb(4\pi\rho)^{- 1/2}$, K\'arm\'an-Taylor constants $\alpha$ and $\beta$ \citep{matthaeus1996jpp,breech2008turbulence},
and functions of cross helicity $f^{\pm}(\sigma_c) = (1 -
\sigma_c^2)^{1/2}[(1 + \sigma_c)^{1/2} \pm (1 - \sigma_c)^{1/2}]/2$ which account for the effect of dynamical alignment
\citep{matthaeus2004grl}. 
The right-hand side of Equation \eqref{z2eq} is the von Karman turbulence heating rate \citep{karman1938prsl} adapted for MHD \citep{hossain1995PhFl,wan2012JFM697,bandyopadhyay2018prx} and plasma \citep{wu2013prl}. The fluctuation energy loss due to von Karman decay is balanced in a quasi-steady state by an 
internal energy supply term in the pressure equations (\ref{pseq1}--\ref{peeq1}), 
which takes the form 
\( Q_T = \alpha f^{+}(\sigma_c)\rho Z^3/(2\lambda)\). 
The following key assumptions were made in deriving Equations (\ref{z2eq}--\ref{lameq}):
local incompressibility of fluctuations, which are also assumed to be transverse to, and axisymmetric about, the mean field; constant normalized difference energy $\sigma_D$ [entering Equation (\ref{z2eq}) as a parameter]; and a single correlation length $\lambda$ \citep[cf.][]{Zank2017ApJ835}. \added{The latter assumption implies structural similarity of autocorrelation functions of the turbulent fields \(\bm{b}',~\bm{v}',\) and \(\bm{z}_\pm,\), and therefore that the similarity/correlation scales of each of these fields are identical. See Appendix \ref{sec:app} for a test of this assumption using PSP data.} Notably, we set the polytropic index to the adiabatic value $\gamma = 5/3$, so that solar wind heating emerges solely from the turbulent cascade and the divergence of electron heat flux.
\subsection{Model Parameters and Numerical Implementation} \label{sec:param}
We solve the mean-flow equations together with the turbulence transport equations in the spherical shell between the coronal base (just above the transition region) and the heliocentric distance of 5 au; the computational domain is divided into two regions: the inner (coronal) region of \(1 - 30~\rs\) and the outer (solar wind) region between \(30~\rs - 5~\text{au}\). The relaxation method, i.e., the integration of the (time-dependent) equations in time until a steady state is achieved, is used in both regions. The simulations have a resolution of \(702\times 120\times 240\) grid points along \(r\times \theta\times \phi\) coordinates. The computational grid has logarithmic spacing along the heliocentric radial (\(r\)) direction, with the grid spacing becoming larger as \(r\) increases. The latitudinal (\(\theta\)) and longitudinal (\(\phi\)) grids have equidistant spacing, with a resolution of 1.5\degree~each. In terms of physical scales, the grid spacing corresponds to several correlation lengths of magnetic fluctuations \citep[e.g.,][]{Ruiz2014SoPh}, thus providing strong motivation for the 
statistical model we employ for unresolved, subgridscale turbulence.

Boundary conditions are specified at the coronal base using Wilcox Solar Observatory (WSO) and ADAPT magnetograms \citep[which are based on the GONG magnetogram;][]{arge2010AIPC}. WSO and ADAPT magnetograms are scaled by a multiplicative factor of 8 and 2, respectively,\footnote{This scaling is required to obtain agreement between model results and spacecraft observations near Earth \citep[see][]{riley2014SoPh}. The choice of scaling factor and its effects on the model output are discussed in detail by \cite{usmanov2018}.} and smoothed using a spherical harmonic expansion up to \(9^\text{th}\) and \(15^\text{th}\) order, respectively. Input parameters specified at the coronal base include: the driving amplitude of Alfv\'en waves (30 km~s$^{-1}$), the
density ($8 \times 10^7$ particles cm$^{-3}$), the correlation scale of turbulence (\(10,500\)~km), and temperature ($1.8 \times 10^6$~K). The cross helicity in the initial state is set as \(\sigma_c = -\sigma_{c0} B_r/B_r^\text{max}\), where  \(\sigma_{c0}=0.8\), \(B_r\) is the radial magnetic field, and \(B_r^\text{max}\) is the maximum absolute value of \(B_r\) on the inner boundary. The input parameters also include the fraction of turbulent energy absorbed by protons $f_p = 0.6$, energy difference \(\sigma_D=-1/3\), and  \added{K\'arm\'an-Taylor constants \(\alpha = 2\beta = 0.128\)}. Further details on the numerical approach and initial and boundary conditions may be found in \cite{usmanov2018}, \added{who also examined the influence of varying these parameters on the model results}. 

The model is well-tested and has been shown to yield reasonable agreement with observations \citep{breech2008turbulence,usmanov2011solar,usmanov2012three,chhiber2017ApJS230,usmanov2018,chhiber2018apjl,chhiber2019psp1,bandyopadhyay2020ApJS_cascade,ruffolo2020ApJ}. For the present study we performed ten runs using both WSO and ADAPT magnetograms corresponding to each of the five PSP solar encounters considered. In the following we show results based on the magnetogram that yielded the best agreement with PSP data for each orbit. The runs shown are: (I) ADAPT map with central meridian time 2018 November 6 at 12:00 UTC for orbit 1; (II) WSO map for Carrington Rotation (CR) 2215, for orbit 2; (III) WSO map CR 2221 for orbit 3; (IV) WSO map CR 2226 for orbit 4; (V) WSO map CR 2231 for orbit 5. \added{The ADAPT maps have a 1\degree-resolution both in heliolatitude and heliolongitude. The synoptic magnetograms from the WSO have 5\degree\ resolution in heliolongitude and 30 points equidistantly distributed over the sine of heliolatitude.}
A detailed examination of differences between simulations based on magnetograms from different observatories is left for future work. 

\section{Results} \label{sec:PSP}
\subsection{PSP Data}\label{sec:data}
\begin{figure}
    \centering
    \includegraphics[width=.45\textwidth]{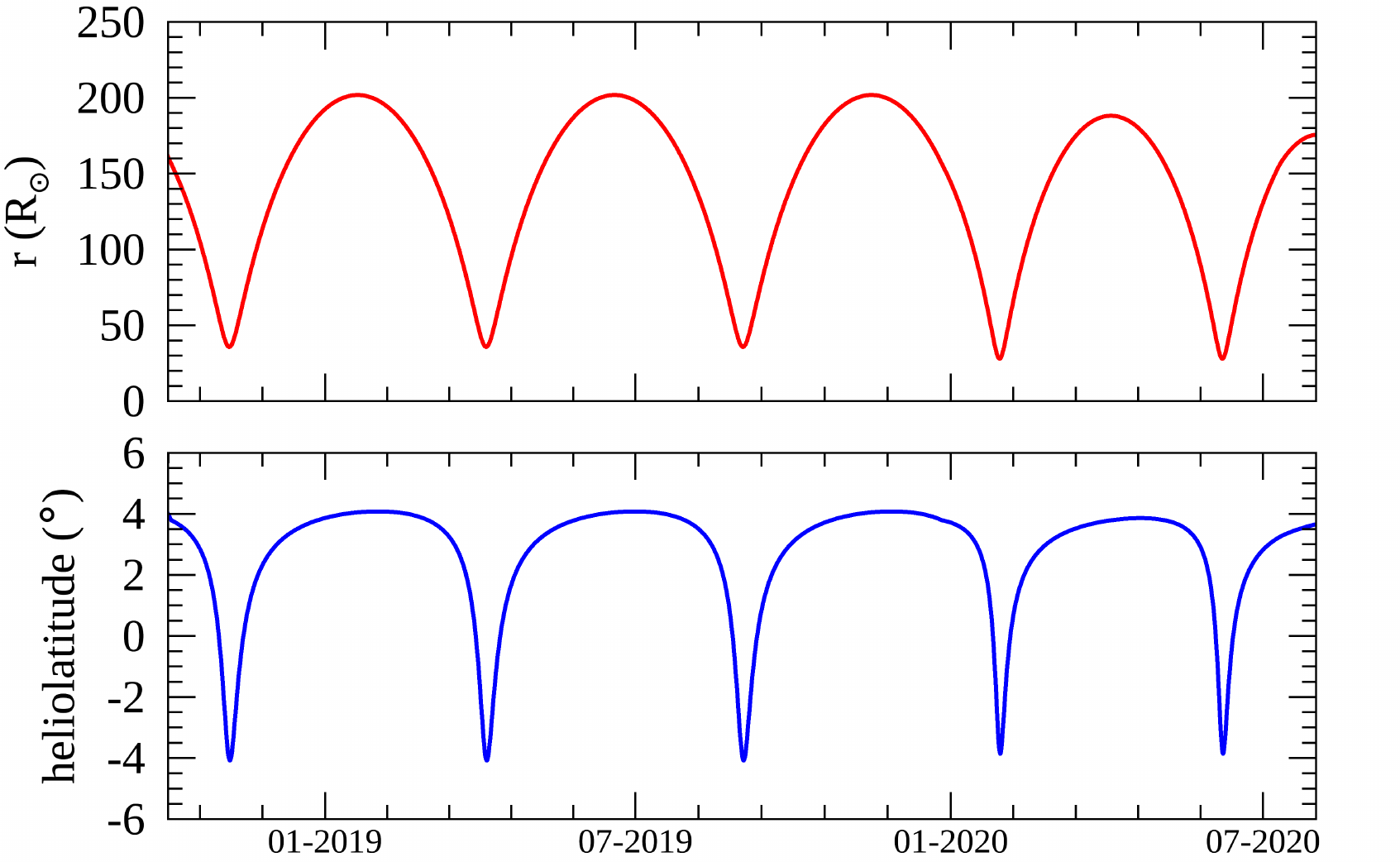}
    \caption{PSP's heliocentric position and heliolatitude during its first five orbits. Times shown span UTC 2018 Oct 01 to 2020 Aug 01.}
    \label{fig:position}
\end{figure}

We use publicly available data for the first five orbits of PSP covering the period between October 2018 to July 2020. Magnetic field data are from the fluxgate magnetometer (MAG), part of the FIELDS instrument suite \citep{bale2016SSR}, and plasma data are from the Solar Probe Cup (SPC) on the SWEAP suite \citep{kasper2016SSR,case2020ApJS}. Level 2 MAG data and Level 3 SPC moment data \citep[see][]{case2020ApJS} are resampled to 1-s cadence using a linear interpolation. SPC data are cleaned using a Hampel filter in the time domain \citep{pearson2002hampel,bandyopadhyay2018filter,parashar2020ApJS}. 
PSP's heliocentric position and heliolatitude at the time of these measurements are shown in Figure \ref{fig:position}. Perihelia are at \(35.6\ \rs\) for orbits 1 to 3, and at \(\sim 28\ \rs\) for orbits 4 and 5. PSP stays close to the ecliptic plane in its highly elliptical orbit \citep{fox2016SSR}.

\subsection{Comparisons of Time Series}\label{sec:tser}
We use trilinear interpolation to obtain model results along the PSP trajectory (at 1-min cadence). To compare PSP data with mean (resolved) fields from the model (see Section \ref{sec:2fluid}) 
we coarse-grain PSP observational data so they correspond roughly to the mean variables in the Reynolds-averaging procedure. To accomplish this we smooth the 1-s cadence PSP data using a boxcar average over a moving window of 2-hour duration\footnote{The selection of an averaging interval can influence the results of these types of analyses. Several previous papers document the kinds of effects that are expected. \cite{isaacs2015JGR120} showed general effects of averaging interval choice in 1 au data. For PSP, \cite{parashar2020ApJS} showed effects of averaging on determination of correlation lengths and densities. Figures 6 and 7 of \cite{ruffolo2020ApJ} show how choice of averaging interval suppresses fluctuations but does not appear to change the temporal/radial trend of quantities such as Alfv\'en speed. The window of 2 hours chosen here is selected to be ``safely'' in the range of several correlation scales, to emphasize trends at the outer scale of the turbulence, rather than variations within the inertial range.}, and then downsample the data to 1-hour cadence. Unless otherwise indicated, in the following \(\langle \dots\rangle\) refers to this boxcar average which, in principle, when applied to the PSP fields, produces 
a coarse-graining that 
corresponds to the 
mean fields explicitly resolved in the Reynolds-averaged simulations. Note that the grid resolution of the simulations corresponds to a temporal scale of about 1 hour, assuming a solar wind speed of 300 km/s (roughly the average value during PSP's solar encounters; see below). This procedure is applied to proton velocity \(\bm{v}\), density \(n_p\), and thermal speed \(\omega_p\), as well as the magnetic field \(\tilde{\bm{B}}\). The proton temperature is then \(T_p = m_p \langle \omega_p^2\rangle/k_\text{B}\).

To compare the observations with turbulence parameters from the model (see Section \ref{sec:trans}) we compute velocity and magnetic fluctuations from the 1-s cadence PSP data as follows. The magnetic field \(\tilde{\bm{B}}\) is converted to Alfv\'en units using the formula \(\bm{b} = \tilde{\bm{B}}/\sqrt{4\pi\langle\rho\rangle}\), where \(\langle\rho\rangle\) is the boxcar-averaged proton mass density. Velocity and magnetic fluctuations are then computed as \(\bm{v}' = \bm{v} - \langle\bm{v}\rangle\) and \(\bm{b}' = \bm{b} - \langle\bm{b}\rangle\). From these, Els\"asser variables are computed as \(\bm{z}_\pm = \bm{v}' \pm \bm{b}'\) \citep{elsasser1950PhRv}. The average turbulence energy is then \(Z^2 = \langle v'^2\rangle + \langle b'^2\rangle\), and the normalized cross helicity is \(\sigma_c = (Z_+^2 - Z_-^2)/(Z_+^2 + Z_-^2)\), where \(Z_\pm^2 = \langle z_\pm^2\rangle\). Both \(Z^2\) and \(\sigma_c\) are downsampled to 1-hour cadence. 

\added{As described in Section \ref{sec:trans} and Appendix \ref{sec:app}, the assumption of a single correlation scale implies that \(\lambda\) from the model can be associated with any of the turbulent fields \(\bm{b}',~\bm{v}',\) and \(\bm{z}_\pm,\). See Appendix \ref{sec:app} for a test of this assumption using PSP data. We choose to compare the modeled \(\lambda\) with the observed magnetic correlation length since magnetic field data from PSP are higher quality than plasma data, especially outside of the solar-encounter periods.} The correlation time of magnetic fluctuations \(\tau_c\) is computed as the time lag at which the Blackman-Tukey autocorrelation function \citep{matthaeus1982JGR} falls to \(1/e\) of its value at zero lag. 
The averaging for computing the autocorrelation is performed over 24-hr intervals, thus yielding a daily value for \(\tau_c\). Assuming the validity of the Taylor hypothesis
and using the daily mean of the solar wind speed, the correlation time is converted to a correlation length \(\lambda\). Note that the Taylor hypothesis has been shown to be reasonably valid during these first PSP perihelia \citep{chhiber2019psp2,chhiber2020ApJS,chen2020ApJS,perez2021AA}.

\begin{figure} 
\centering
		\includegraphics[width=.49\textwidth]{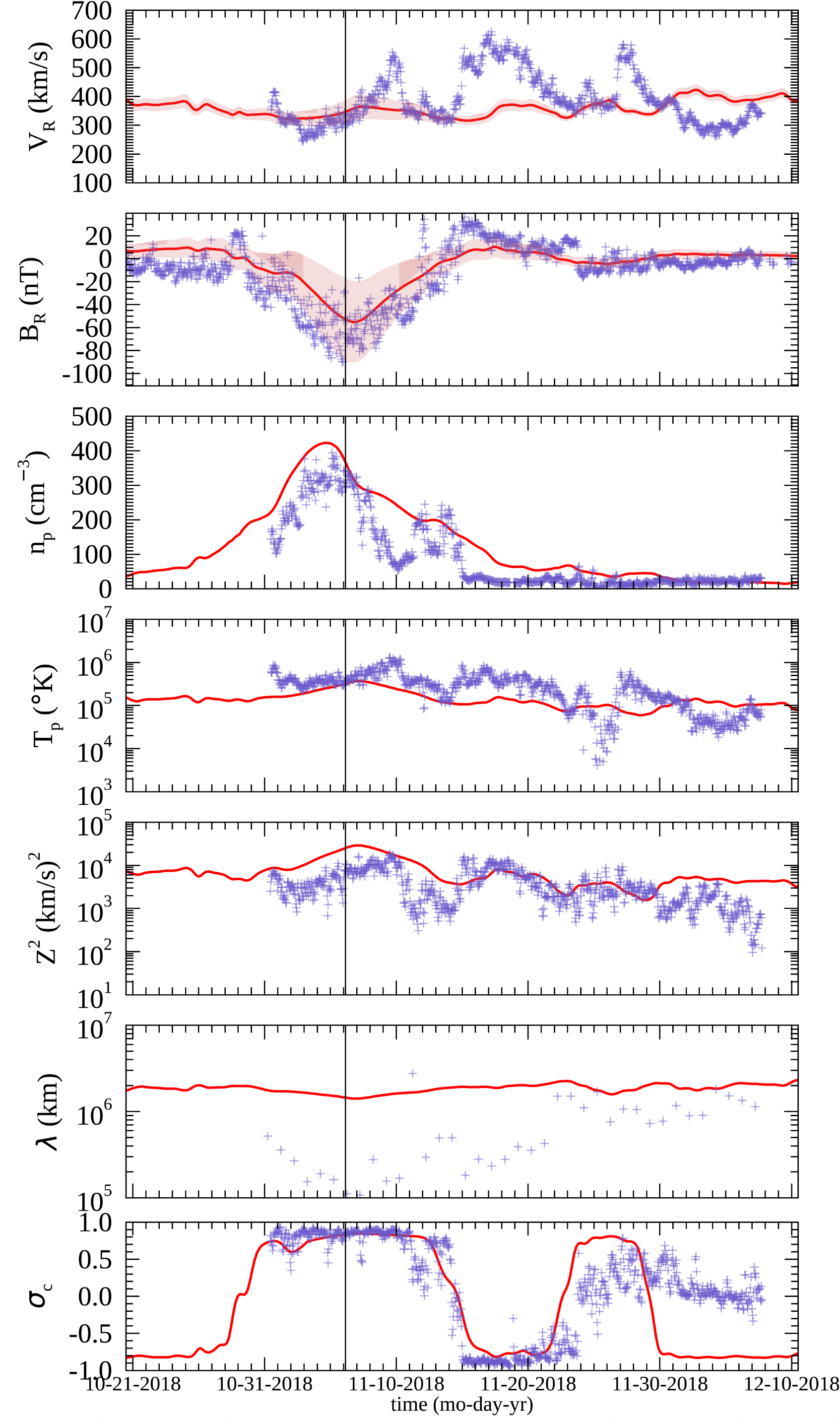}
		\caption{Blue `+' symbols show PSP data from orbit 1, plotted at 1-hour cadence except \(\lambda\), for which daily values are shown. Red curve shows results from the model, sampled along a synthetic PSP trajectory. Quantities shown are mean radial velocity of ions (\(V_R\)), mean radial magnetic field \(B_R\), mean ion density \(n_p\), mean ion temperature \(T_p\), mean turbulence energy \(Z^2\), correlation length of magnetic fluctuations \(\lambda\), and normalized cross helicity \(\sigma_c\). The shading in the top four panels marks an envelope obtained by adding and subtracting the local turbulence amplitude from the model to the mean value from the model (see the text for details). The vertical black line marks perihelion. The model uses ADAPT map with central meridian time 2018 November 6 at 12:00 UTC (Run I). Minor ticks on the time axis correspond to 1 day.} 
		\label{fig:O1}
\end{figure}

\begin{figure} 
\centering
		\includegraphics[width=.49\textwidth]{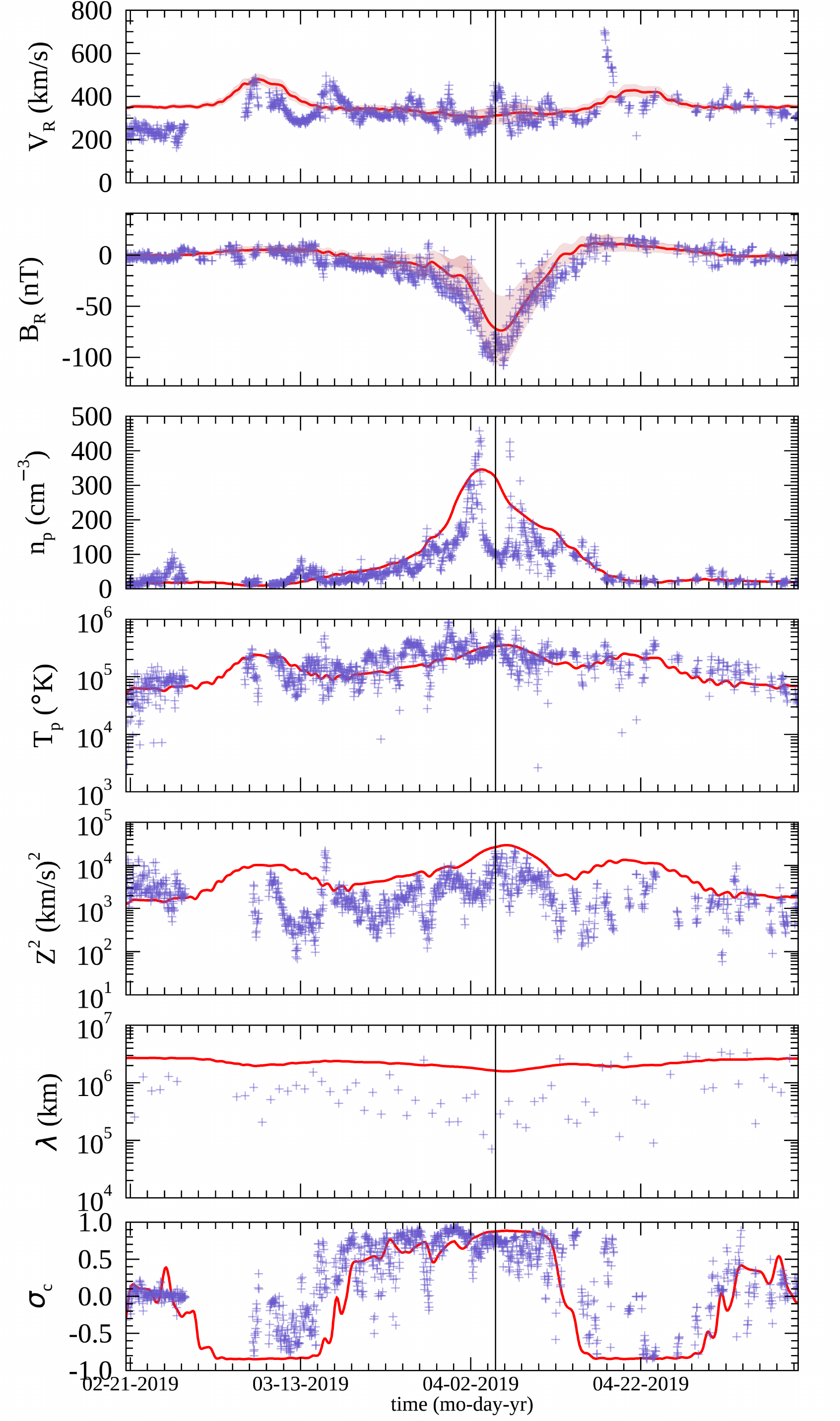}
		\caption{PSP orbit 2 Data in comparison with global simulation based on WSO map for CR 2215 (Run II). The description follows Figure \ref{fig:O1}. Minor ticks on the time axis correspond to 2 days.}
		\label{fig:O2}
\end{figure}

\begin{figure} 
\centering
		\includegraphics[width=.49\textwidth]{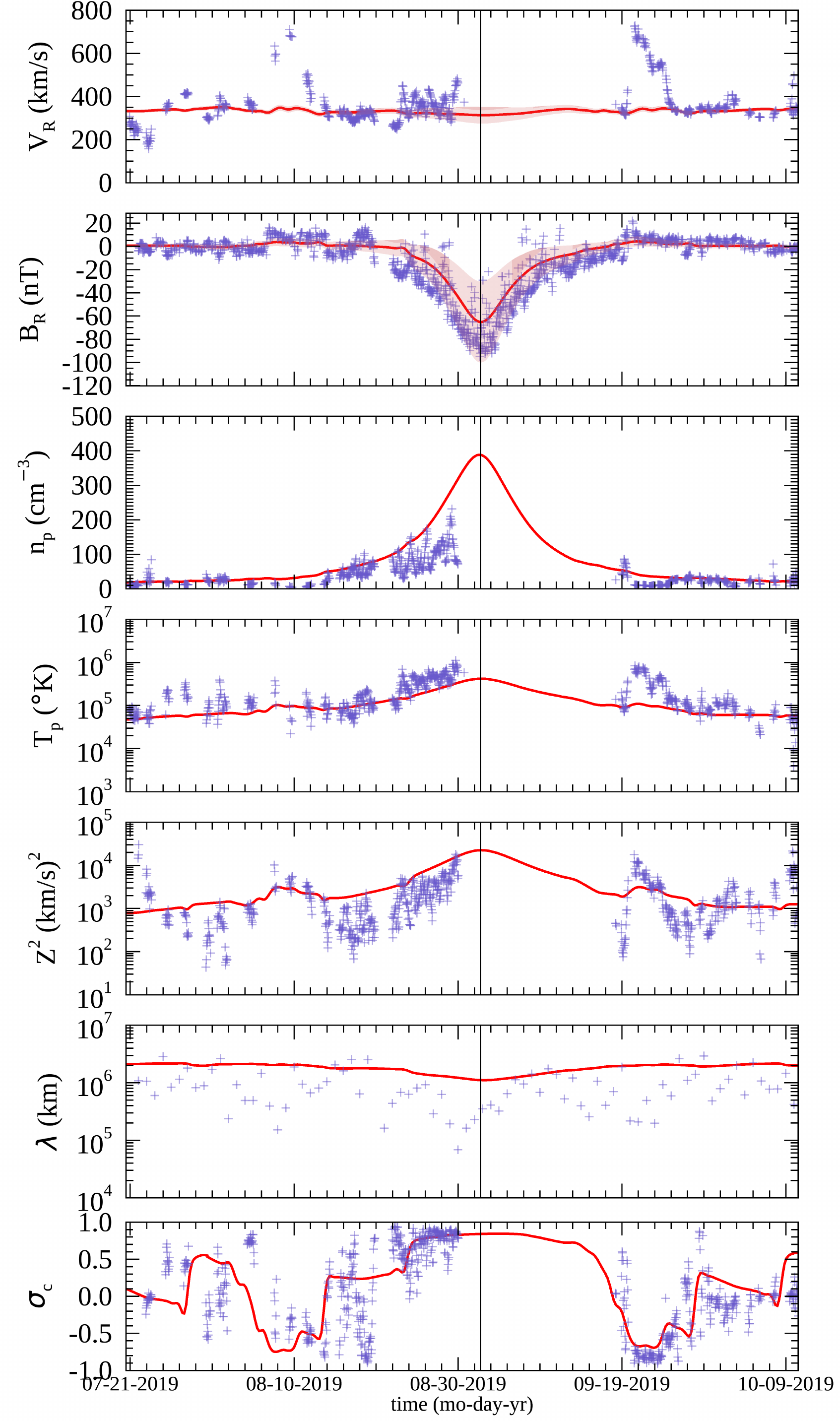}
		\caption{PSP orbit 3 Data in comparison with global simulation based on WSO map for CR 2221 (Run III). The description follows Figure \ref{fig:O1}. Minor ticks on the time axis correspond to 2 days.}
		\label{fig:O3}
\end{figure}

\begin{figure} 
		\includegraphics[width=.49\textwidth]{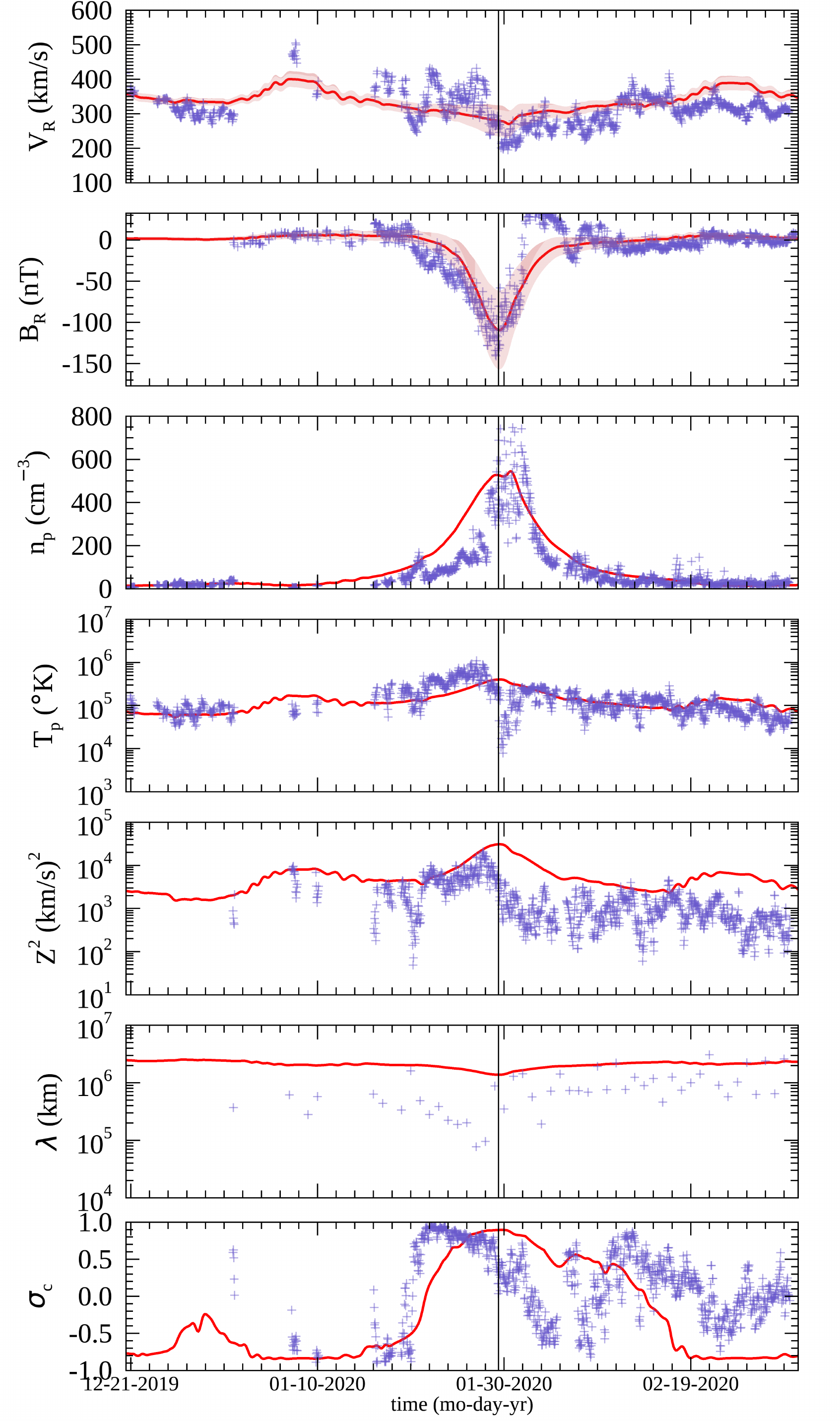}
		\caption{PSP  orbit 4 Data in comparison with global simulation based on the WSO map for CR 2226 (Run IV). The description follows Figure \ref{fig:O1}. Minor ticks on the time axis correspond to 2 days}
		\label{fig:O4}
\end{figure}

\begin{figure} 
		\includegraphics[width=.49\textwidth]{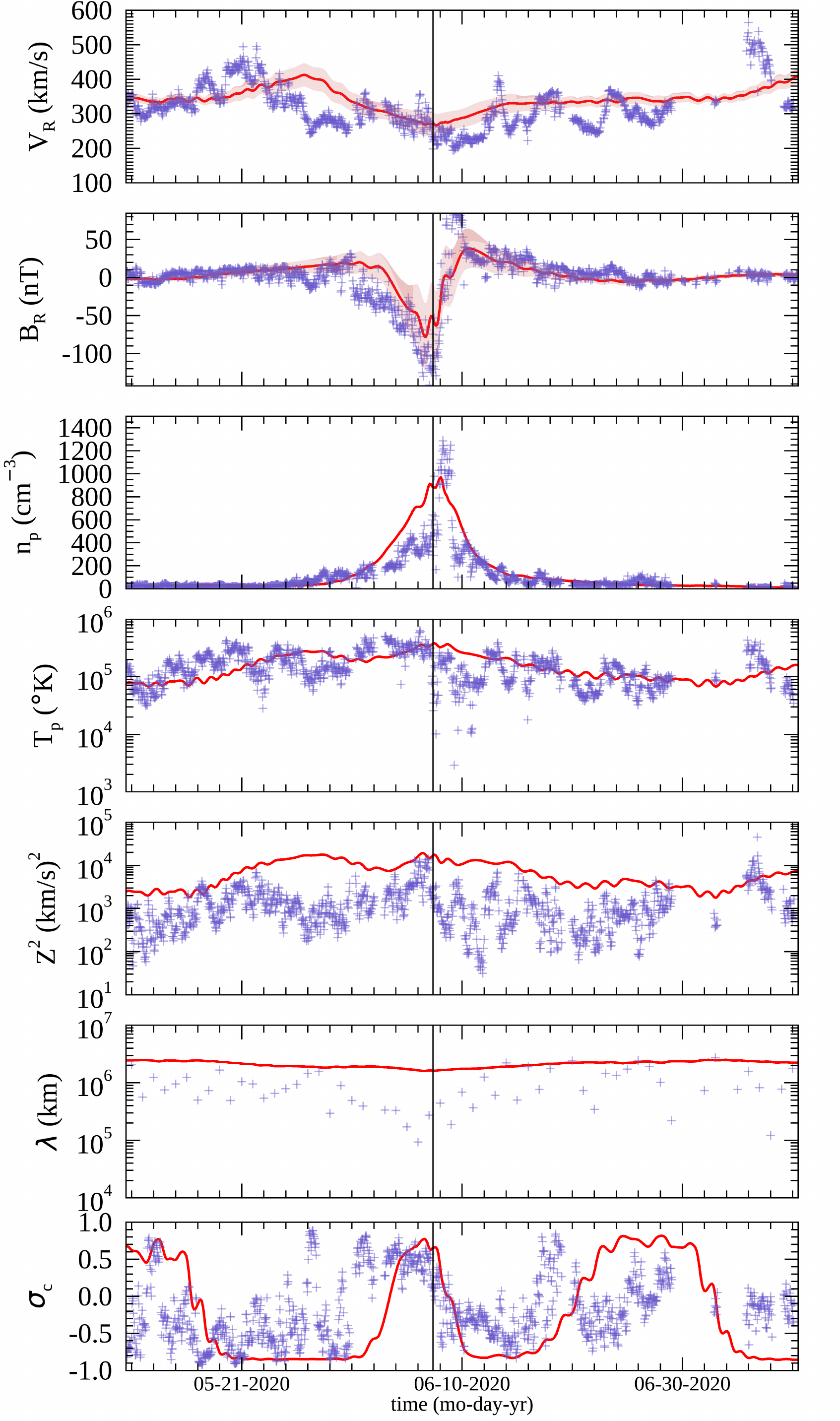}
		\caption{PSP orbit 5 data in comparison with global simulation based on WSO map for CR 2231 (Run V). The description follows Figure \ref{fig:O1}. Minor ticks on the time axis correspond to 2 days.}
		\label{fig:O5}
\end{figure}

Figures \ref{fig:O1} to \ref{fig:O5} 
show comparisons between PSP data from the first five orbits with 
global simulation data.
\footnote{We do not show T and N components \citep[in a heliocentric RTN coordinate system; e.g.,][]{franz2002pss} of ion-velocity and magnetic fields in the time-series comparisons in the present subsection. On average, \(V_N\) and \(B_N\) tend to stay close to zero throughout an orbit, both in the model and in observations. The tangential velocity \(V_T\) shows a systematic increase near perihelia in the observations \citep{kasper2019Nat}, which is not captured by the model \cite[see also][]{reville2020ApJS}. The observations also show a slight increase in \(B_T\) near perihelia, and this is captured well by the model. Both \(V_T\) and \(B_T\) \textit{are} shown in Section \ref{sec:radial} (see Figure \ref{fig:radial}). See also Figure 4 of \cite{ruffolo2020ApJ} for a time-series comparison of T and N components of ion-velocity and magnetic fields between PSP's first orbit and the model.} In these figures the plots of the radial components of ion velocity $V_R$ and the magnetic field $B_R$ are supplemented by a shaded region that is obtained from turbulence energy $Z^2$ in the simulation. This represents turbulent fluctuations that are not explicitly resolved. \(Z^2\) can be used to obtain estimates for turbulent magnetic and velocity energies by assuming an appropriate value for the Alfv\'en ratio \(r_\text{A} = \langle v'^2\rangle/\langle b'^2\rangle \). Following observations during the PSP encounters \citep{chen2020ApJS,parashar2020ApJS} we assume \(r_\text{A}=0.5\), which is also consistent with the constant value of energy difference \(\sigma_D=-1/3\) assumed in our model (Section \ref{sec:param}). Then the magnetic fluctuation energy becomes \(\langle B'^2\rangle = Z^2 4\pi\rho/(r_\text{A}+1)\).

An analogous procedure is used to obtain the partitioning of fluctuation energy among the three cartesian components of polarization. The transport model in Sec. \ref{sec:trans} adopts the approximation that the fluctuations have purely transverse polarizations. However, it is well known that incompressive, Alfv\'enic-type fluctuations must, at finite amplitude, include three components of polarization. Following observations \citep{belcher1971JGR} we adopt a partitioning in which the RTN components of magnetic fluctuations have variances in a 5:4:1 ratio \citep{belcher1971JGR}, where we associate the `5' and `4' factors with the N and T components, respectively. Therefore radial fluctuations have a variance \(\langle B_R^{\prime 2}\rangle = \langle B^{\prime 2}\rangle/10\). This allows us to generate an envelope around the resolved field $B_R(t)$ that spans the range $B_R (t) - \sqrt{\langle B_R^{\prime 2}\rangle} (t)$ to $B_R (t) + \sqrt{\langle B_R^{\prime 2}\rangle} (t)$. A similar procedure is used to generate an envelope around \(V_R\), except that we assumed the RTN components had variances in a ratio 2:2:1, where the `1' factor is associated with the R component \citep{oughton2015philtran}.

Before proceeding with a detailed discussion of the results, we note that PSP data have relatively larger gaps and lower cadence far from the encounter periods (roughly ten days about perihelia), especially for plasma measurements \citep{case2020ApJS}. With that caveat stated, we move on to Figure \ref{fig:O1}, where comparisons for the first orbit are illustrated. 
For the radial velocity 
we see no detailed agreement of PSP and the simulated 
values. Even the envelope of fluctuations fails to 
encompass the variations in the PSP values of $V_R$. Note that the first encounter was somewhat atypical in that PSP observed fast wind emanating from a low-latitude coronal hole \citep{badman2020ApJS} and this was not captured in the model, possibly due to low resolution of the magnetogram data. The situation is considerably better for the 
comparison of values of $B_R$. The resolved simulation values have a very similar shape to the PSP data, although the magnitudes are systematically low. However, the shaded region provides an envelope that neatly spans the range of observed PSP values near perihelion (marked with a vertical solid line). The third panel compares the proton densities. The general shapes of simulated and PSP values are similar, although the rapid variations of the observed values are not captured by the simulated data. Note the dip in density that occurs simultaneously with the spike in velocity near 11-10-2018, once again signifying connectivity to that equatorial coronal hole. Note also that no density fluctuation amplitude is available in our subgrid formulation, which assumes small-scale incompressibility. The panel showing the proton temperatures again shows only modest agreement. Note that plasma data from PSP were not available in the first week of this period, and the associated PSP curves are empty in that range in all panels except $B_R$.

The panel that compares subgrid turbulence energies $Z^2$ indicates somewhat better comparison with the levels and some of the modulated features of the PSP observations are accounted for reasonably well in the simulated data. The next panel, the correlation scale $\lambda$, does not compare at all well, with the simulated value being an order of magnitude larger than the same quantity computed from PSP data near the encounter (see discussion in Section \ref{sec:radial}). Finally, the  normalized cross helicity comparison gives mixed results, showing very good agreement near perihelion, and accurate transition to negative values around 11-14-2018. Note that this transition indicates a crossing of the heliospheric current sheet (HCS) by PSP into a region of opposite magnetic polarity \citep{szabo2020ApJS}. The subsequent return to positive $\sigma_c$ around 11-26-2018 reaches reasonable values, but with a time shift between PSP and the simulation results. Finally, in the days leading to 12-10-2018, the simulation fails altogether (see discussion in Section \ref{sec:radial}). 

Figures \ref{fig:O2} through \ref{fig:O5}
are presented in the same format as Figure \ref{fig:O1}
and contain the same comparisons of data from simulation 
and PSP observation, in these cases for orbits 2 through 5, respectively. 
%

Figure \ref{fig:O2} shows orbit-2 simulation vs. PSP comparisons that are somewhat better than those shown for orbit 1 especially for \(V_R\) and \(T_p\). Large transients seen in the radial velocity in Figure \ref{fig:O1} are essentially absent in Figure \ref{fig:O2}. We also note that \(Z^2\) from the model tends to skirt the upper envelope of the PSP data.   

Plasma measurements during the third PSP encounter were not available and measurements outside the encounter are somewhat patchy, as seen in Figure \ref{fig:O3}. Transient features seen in \(V_R\) are once again not captured by the simulation, which shows a very steady radial wind speed of about 340 km/s. However, the radial magnetic field is well represented. We remind the reader that the correlation lengths $\lambda$ plotted in these comparisons are the magnetic correlation lengths. There is somewhat better agreement between the model and observations for this quantity compared with the first two orbits. We also note the presence of relatively high cadence changes in $\sigma_c$ during the third orbit,\footnote{More frequent crossings of the HCS are the likely reason, although the patchiness of the SPC data during this orbit may also be contributing.} some of which are captured well by the model.

In Figure \ref{fig:O4} one sees a set of comparisons for orbit 4 that fit the prior pattern, with transient $V_R$ periods in the PSP data not well represented in the simulation, and in contrast, a better comparison with radial magnetic field, although some HCS crossings are missed. This latter effect is evident from the cross-helicity comparison as well. Density, temperature, and turbulence level display reasonable agreement, with some systematic time shifting of the the peaks relative to the maxima of the simulation data. As with the other orbits, the simulation correlation length is systematically larger than that measured by PSP. 

The comparisons in Figure \ref{fig:O5} for orbit 5 
are consistent with those seen previously. The asymmetry in \(B_R\) across the perihelion is captured well by the model. 
The quality of the comparison of cross helicities 
is rather mediocre for this orbit, while the $Z^2$ values are systematically above the PSP observations, as are the correlation scales.

In the next section we discuss the overall trends that can be inferred from aggregating data over five orbits, as well as possible reasons for some of the discrepancies between the model and the observations.
\subsection{Radial Trends Aggregated from Five Orbits \label{sec:radial}}
\begin{figure*}[ht]
    \centering
    \includegraphics[width=.9\textwidth]{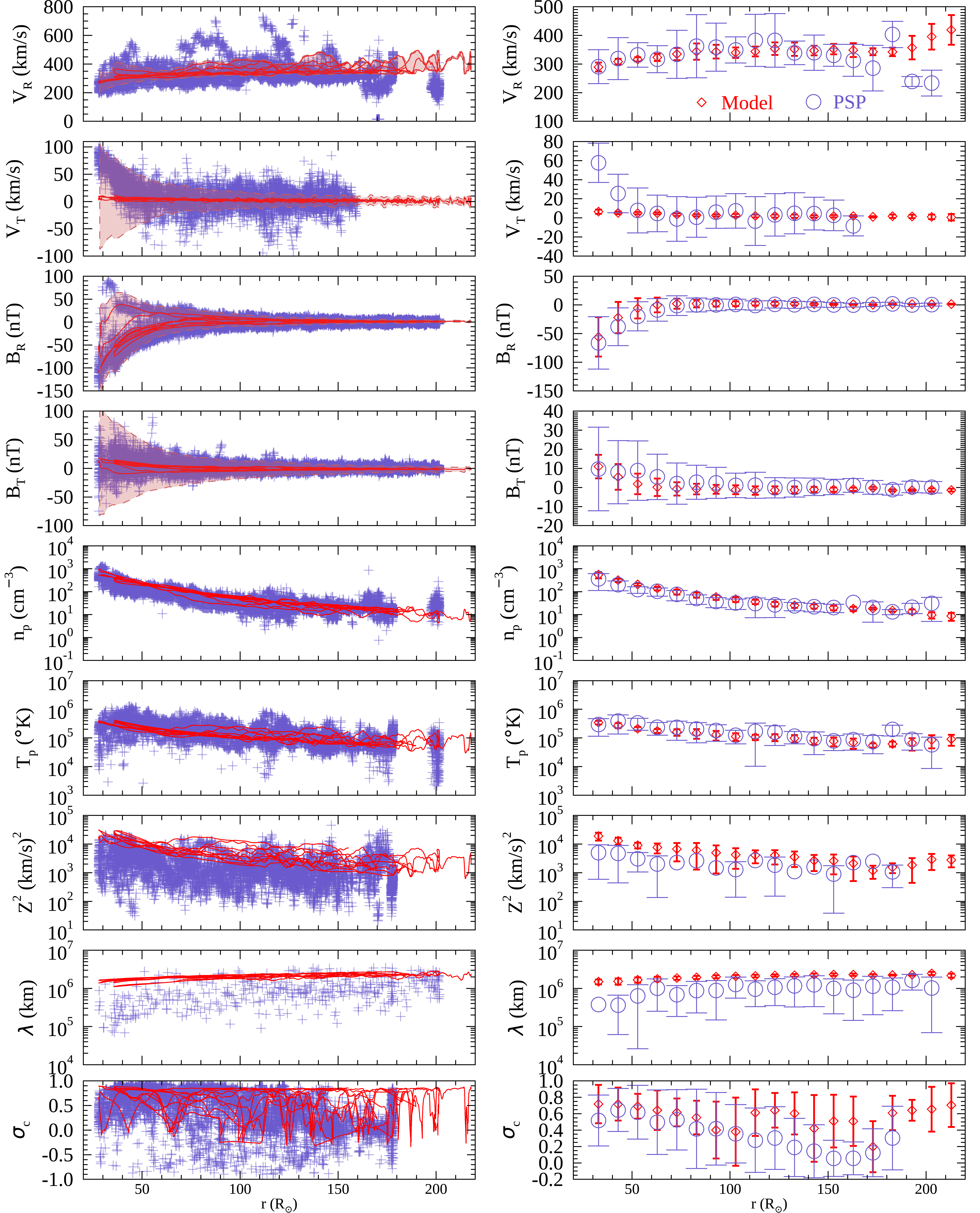}
    \caption{\textit{Left}: Model results (red curve) compared with PSP data (blue `+' symbols) from the first five orbits. Top two panels include shaded regions representing fluctuation amplitudes derived from the model (see text). \textit{Right}: Mean values within bins of \(10~\rs\) from the model (red diamonds) and PSP data (blue circles). Bars above and below symbols represent the standard deviation; Bars that extend to negative values are not shown on logarithmic axes. \(\sigma_c\) has been sector rectified (see text).}
    \label{fig:radial}
\end{figure*}
A different approach to comparing PSP and simulation data is to look at the radial trends 
in each, in the sense of a superposed radial epoch. 
By combining data from different PSP orbits and employing 
simulation data based on the corresponding magnetograms for 
separate orbits, we can understand better the behavior of the mean radial trends as well
as the influence of transients. 

The left column of Figure \ref{fig:radial} shows data aggregated from the first five orbits as a function of heliocentric distance. Also shown are results from simulated PSP trajectories through five simulations corresponding to the five orbits (Runs I-V). Note that the five orbits yield five envelopes each for each of \(V_R, V_T, B_R\), and \(B_N\), taking into account the rms fluctuation level from the model as described in Section \ref{sec:tser}. For visualization purposes the envelope shown in Figure \ref{fig:radial} is generated by picking the maximum and minimum values from the envelopes for the five orbits, within radial bins of size \(1~\rs\). The cross helicity has been sector rectified so that positive \(\sigma_c\) indicates outward (away from Sun) propagation of Alfv\'en waves \citep[e.g.,][]{barnes1979inbook,roberts1987JGRb}; i.e., when the mean radial magnetic field (at 1-hour cadence) has positive polarity the cross helicity is multiplied by \(-1\): for all time \(t\) when \(B_R(t) \geq 0\) we let \(\sigma_c (t) \to  -\sigma_c(t)\).

To obtain a clearer picture of radial trends, we compute mean values and standard deviations within radial bins of size \(10~\rs\) for each quantity shown in the left column of Figure \ref{fig:radial}. This is done separately for PSP data and the model, and the results are shown in the right column of Figure \ref{fig:radial}; symbols show mean values and vertical bars represent one standard deviation above and below the mean. \added{Our goal here is to examine long-term radial trends in the ecliptic region during solar-minimum conditions, as observed by PSP during its first five orbits. In the hourly-cadence PSP data presented in the left column of Figure \ref{fig:radial}, only 1.7\% of points have \(V_R > 500\) km/s; 2.9\% have \(V_R > 450\) km/s, and 5.5\% have \(V_R > 400\) km/s; therefore the data set is overwhelmingly representative of slow wind conditions. Note that fast wind is often described as having \(V_R > 500\) \citep[e.g.,][]{dasso2005ApJ}.}

The radial trend of mean 
radial velocities in Figure \ref{fig:radial}
is very similar in the PSP data and the simulations. 
The spread in radial velocities at each position, 
indicated by bars above and below the symbols, is much greater for the PSP data, consistent with the discussion above for Figures \ref{fig:O1}-\ref{fig:O5}. We see from the second panel of the figure that the strong tangential flows apparently observed near perihelia \citep{kasper2019Nat} are not reproduced in the modeled \(V_T\).\footnote{We do not show PSP observations of \(V_T\) above \(160~\rs\) since the data are extremely noisy.} Note that the envelope of modeled fluctuations (left column) is symmetric about the mean \(V_T\), and therefore cannot account for the systematic increase in the observed mean \(V_T\) near perihelia. The reasons behind this discrepancy are not currently understood and lie beyond the scope of the present study \citep[cf.][]{reville2020ApJS}.
The comparisons of radial trends in $B_R$, \(B_T\), $n_p$, and $T_p$ are much tighter, as is evident in the very good coincidence of the mean values 
from simulation and from PSP in these three quantities. The relatively small 
standard-deviation bars in \(B_R, n_p,\) and \(T_p\) reflect the smoothness of
these quantities in the averaged PSP data and also 
demonstrate the accuracy of the global MHD simulation.

The comparisons of the turbulence parameter $Z^2$ are reasonable; however, the modeled turbulence energy clearly skirts the upper envelope of the observations (left column), with the mean value from the model generally staying a factor of 1.5-2 larger (right column). The correlation scale is slightly larger in the model for most radial distances, but below \(\sim 60~\rs\) the discrepancy widens. \added{A possible reason could be that PSP samples slab-like fluctuations at an increasing rate near perihelia, as the magnetic field becomes more radial; in contrast, the correlation length in the model is associated with quasi-2D fluctuations, identified with a turbulent cascade perpendicular to the mean field \citep[][]{oughton2015philtran}. We have checked that the fraction of points with \(\bm{B}\) lying within 30\degree~of the ion-velocity direction (quasi-alignment of \(\bm{B}\) and \(\bm{V}\)) increases from \(\sim 0.15\) to \(\sim 0.6\) as one moves from \(175~\rs\) to \(30~\rs\). Indeed, \cite{adhikari2020ApJ} use a turbulence transport model that includes separate correlation scales for slab and 2D fluctuations to find that the modeled slab correlation scale better matches the correlation scale observed by PSP, when focusing on intervals characterized by quasi-alignment. These findings are supported by \textit{Helios} observations analyzed by \cite{ruiz2011JGR}, who also found that intervals with smaller turbulence ``age'' \citep{matthaeus1998JGR} are more slab-like, and that the associated correlation scales in parallel intervals are smaller than the correlation scales of perpendicular intervals, as one approaches the sun.}

Finally, as the radial distance decreases the observed sector-rectified cross helicity systematically increases from near zero to \(\sim 0.6\), with large standard deviations \citep[see also][]{chen2020ApJS,parashar2020ApJS}.\footnote{There are instances in the observations when the sector-rectified cross helicity is negative, indicating Sun-ward propagation of Alfv\'en waves; this may be related to the switchback phenomenon \citep{mcmanus2020ApJS}.} The modeled \(\sigma_c\) generally stays at moderate to high values above \(100\ \rs\), while the trend below this radius appears to match observations better. The dip near \(100\ \rs\) followed by the increase toward perihelia is likely due to the ``virtual'' PSP crossing the (mainly equatorial) HCS in the model, which is characterized by low \(\sigma_c\), and subsequently moving into the opposite polarity hemisphere \citep[see][]{chhiber2019psp2}. The large cross helicity in the model at near-Earth distances is likely due to the absence of fine-scale structure \replaced{to drive shear}{which would drive shear and produce gradients in magnetic, density, and velocity fields, any of which can} in turn reduce \(\sigma_c\) \citep{roberts1992jgr,zank1996evolution,breech2008turbulence}.

It is clear from Figure \ref{fig:radial} that the resolved quantities from the model -- \(V_R\), $B_R$, \(B_T\), $n_p$, and $T_p$ -- agree better with the observations compared with the turbulence parameters. In particular, it is interesting that the mean temperature profiles compare extremely well while the turbulence quantities, which ultimately drive heating, fall short of the same level of accuracy. This constrains potential directions for improvement of the turbulence model, which performs very well in accounting for the resolved quantities and notably the temperature.    
\subsection{Synthetic Magnetic Fluctuations Constrained by Turbulence Transport Model \label{synthetic} }
The subgridscale model employed
in previous sections to describe
properties of the
turbulence provides only a statistical 
picture but no information about the waveform of the 
fluctuations. Adding an envelope centered on the 
resolved mean fields, such as the radial magnetic and velocity fields in Figures \ref{fig:O1} - \ref{fig:O5}, leads to a suggestion of a possible range of values, 
but stops short of providing the reader with an
explicit representation of the full magnetic (or velocity) signal. However, armed with values of 
the variance $Z^2$, the correlation scale $\lambda$,
and cross helicity \(\sigma_c\),
the turbulence model provides a basis for 
generating consistent random realizations of the 
relevant variables. For simplicity only the variance
is employed in this
(first) exercise of this type. 

To proceed, one must adopt a probability
distribution for the fluctuations, 
using it to generate a realization for comparison
with the observed PSP signal. Such realizations are 
clearly not unique, but they permit exploration 
of how properties of the underlying probability distribution can influence the quality of the comparison with the observations. Interplanetary magnetic fluctuations at 1 AU have near-Gaussian probability distributions in the inertial range \citep[e.g.,][]{padhye2001JGR,bandyopadhyay2020ApJL_curvature}. However, observations of the near-Sun magnetic field by PSP \citep{bale2019Nat} indicate that fluctuations in the radial component do not have symmetric distributions; instead, these are biased towards reversals in the magnetic polarity, and their probability distributions are characterized by a significant skewness (Chhiber 2021, in preparation). Therefore, the random distribution used to generate realizations of synthetic (radial) magnetic fluctuations in near-Sun space must have finite skewness as a property. Here we use a gamma distribution for this purpose, wherein the skewness scales as the inverse square-root of the so-called shape parameter, also known as the order of the distribution \citep[e.g.,][]{thom1958MWR}. Note that we do not make any attempt here to relate the distribution of fluctuations to underlying physical processes; we simply use a distribution that captures \textit{some} properties of the data (see below).

Our procedure for generating the synthetic fluctuations is as follows. Using PSP's second encounter as an example, we begin with the average turbulence energy \(Z^2\) from Run II, interpolated to the PSP trajectory to obtain a time series at 1-minute cadence, which is then converted to an average magnetic fluctuation energy \(\langle B^{\prime 2}\rangle\), assuming an Alfv\'en ratio of 0.5 (as in Section \ref{sec:tser}). We again assume that the RTN components of the magnetic fluctuations have variances in a 5:4:1 ratio \citep{belcher1971JGR}, so that radial fluctuations have a variance \(\langle B_R^{\prime 2}\rangle = \langle B^{\prime 2}\rangle/10\). Next, the mean value of the standard deviation \(\sqrt{\langle B_R^{\prime 2}\rangle}\) is computed within 24-hour bins along the time-series. Within each 24-hour bin a sequence of 1440 random numbers is generated, which follow a gamma distribution of order 6.\footnote{\footnotesize{We use the IDL function {\fontfamily{qcr}\selectfont \href{https://www.l3harrisgeospatial.com/docs/randomu.html}{randumu}}, which is based on the Mersenne Twister algorithm for generating pseudo-random numbers \citep{matsumoto1998ACM}. The choice of order 6 for the gamma function is based on visual inspection of the resulting synthetic time-series.}} 
The mean of each distribution is subtracted, and the distribution is rescaled so that its standard deviation is equal to the the mean \(\sqrt{\langle B_R^{\prime 2}\rangle}\) in the respective bin. Thus we obtain a synthetic time-series of radial magnetic fluctuations \(B'_R\) at 1-min cadence \citep[within the inertial range of turbulence observed by PSP; e.g.,][]{chen2020ApJS}, constrained by the average local fluctuation amplitude given by the turbulence transport model. 

\begin{figure}
    \centering
    \includegraphics[width=.49\textwidth]{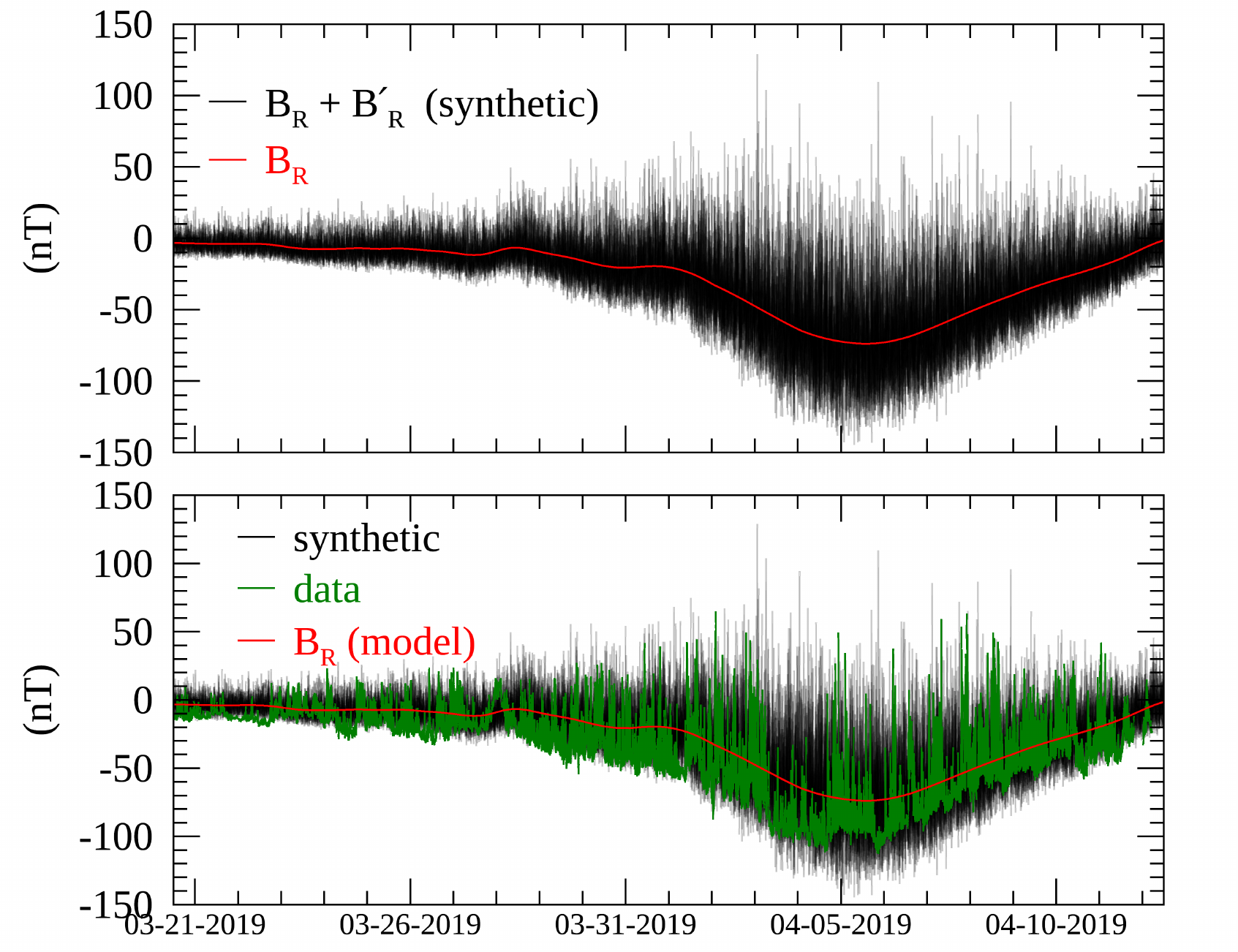}
    \caption{\textit{Top}: Red curve shows mean radial magnetic field along PSP's second orbit, obtained from Run II. Black curve shows a synthetic time series (1-min cadence) of the full radial magnetic field obtained by adding synthetic fluctuations \(B'_R\) to the mean radial magnetic field from the model. These synthetic fluctuations are constrained by the local rms turbulence levels from the model, and are generated using a gamma function distribution of order 6 (see text). \textit{Bottom}: The radial magnetic field observed by PSP is overlaid as a green curve, plotted at 1-min cadence. Black and red curves are the same as in top panel.} \label{fig:synth}
\end{figure}

The top panel of Figure \ref{fig:synth} shows the time series of the radial magnetic field obtained by adding the synthetic fluctuations (\(B'_R\)) to the mean radial magnetic field (\(B_R\)) from the model.
It is evident that the synthetic
signal exhibits a kind of ``one-sided'' 
behavior that is seen in the PSP data. 
This would be lacking in a symmetric synthetic signal (not shown) produced with a non-skewed probability distribution. 
What is lacking in the skewed
synthetic signal, 
as seen in the bottom panel of Figure \ref{fig:synth}, is the \textit{clustering} of large excursions (including switchbacks) 
that is evidently characteristic 
of the PSP signal \citep{chhiber2020ApJS,DudokDeWit2020ApJS}. 

\section{Conclusions and Discussion} \label{sec:discuss}
In this paper we have presented comparisons
of Parker Solar Probe data and global heliospheric MHD 
simulations, employing several novel approaches.
We have employed data and simulations
corresponding to five PSP orbits, providing at least a crude view of the variability expected 
from orbit to orbit in each data type. Our analyses
also involved averaging observational and simulation data in similar ways to achieve commensurate average values and variances. This is facilitated by the Reynolds-averaged structure of the simulation code, which also computes parameters describing local statistics of unresolved turbulence using a well studied (but of course, approximate) transport model for MHD turbulence in an inhomogeneous medium. This strategy enables a statistical comparison of the PSP data with the simulations that would not be possible using only the directly resolved simulation variables, due to the inevitable limitation of spatial resolution \citep[e.g.,][]{miesch2015SSR194}.  

The several comparisons shown
lead to insights concerning the PSP observations as well as a characterization of the limitations of the adopted MHD simulation framework.
Direct comparisons of PSP and simulation time series
show that the coarse features of the observations are reasonably well represented in a steady state model driven by magnetograms. The main shortcomings appear to be the MHD code's inability to reproduce transient and finer scale features of the resolved variables. The statistical turbulence parameters are apparently less accurate; in particular, correlation scales near perihelia are systematically overestimated in the turbulence model.

When the data are reduced to radial profiles, a very good correspondence is found in the comparisons. This agreement is clearer when averages and variances are computed in radial bins. Both resolved and turbulence variables agree well in the averaged radial trends, the only significant departures being seen in the tangential velocities and correlation lengths near perihelia, and in the sector-rectified cross helicity near 1 au. While the reasons behind the discrepancy in \(V_T\) are not presently understood, the offset in correlation scales could be due to PSP sampling \deleted{predominantly} slab-like fluctuations at an increasing rate near perihelia, which are not represented in the predominantly 2D turbulence modeled in our simulations \citep[cf.][]{ruiz2011JGR,adhikari2020ApJ}. The discrepancy in \(\sigma_c\) is understood at least partially as due to the absence of fine-scale structure in the model, which, if present, would drive shear \added{and produce gradients in the mean-flow fields, which can in turn} reduce cross helicity \citep{breech2008turbulence}. Overall it seems quite reasonable that averages over a sufficient number of orbits would suppress effects of transients. This would influence both the observations and the simulations; obtaining agreement between the two provides evidence of consistency with the underlying physics in the numerical model.

Finally we carried out an 
exercise of comparing the observations with 
radial magnetic-field time series
generated from the resolved mean  
field supplemented by synthetic fluctuations, the latter being constrained by the local turbulence amplitude from the model.
A key element in obtaining a reasonable level of comparison is the use of a skewed 
probability distribution in the generation of the synthetic
field. With further refinements, this technique may be employed in numerical studies of energetic particle transport \citep[e.g.,][]{moradi2019ApJ,chhiber2021AA}.

These comparisons of PSP data with the Reynolds-averaged MHD model of \cite{usmanov2018} 
are encouraging with respect to
further applications 
that are supported or enhanced by the ability to provide 3D coarse-grained 
context for interpretation of the observations. Future planned research in this direction will continue for subsequent PSP orbits.
Analogous applications are also anticipated 
for ongoing missions that span the entire heliosphere, including Solar Orbiter \citep{muller2013SoPh}, and missions  under development, such as PUNCH
\citep{deforest2019AGU_PUNCH},
 Helioswarm \citep{spence2019AGU_helioswarm}, and IMAP \citep{mccomas2018SSR}.
A point that we wish to emphasize is that a model such as the  present one that incorporates self-consistent turbulence 
modeling permits a broader range of applications, including studies relating to solar energetic particles, particle scattering \citep[e.g.,][]{wiengarten2016ApJ833,guo2016ApJ,chhiber2017ApJS230,chhiber2021ApJ_flrw,chhiber2021AA},
and other applications that require information about fluctuations \citep[e.g.,][]{chhiber2019psp2,reville2020ApJS}. For example, a quantitative examination of the role of turbulent heating in producing the solar wind, a primary objective of the PSP mission, cannot be fully addressed without a self-consistent model that links the large-scale wind to the properties of fluctuations. 

\begin{acknowledgments}
We thank R. Bandyopadhyay for useful discussions. This research was partially supported NASA Heliophysics Supporting Research program (grants 80NSSC18K1210 and 80NSSC18K1648), and by the Parker Solar Probe Mission. And also was partially supported by NASA LWS grant under Award No. 80NSSC20K0377 and a subcontract from New Mexico Consortium. Computing resources supporting this work were provided by the University of Delaware (Caviness cluster) and by the NASA High-End Computing (HEC) Program (awards SMD-17-5880 and SMD-17-1617) through the NASA Advanced Supercomputing Division at Ames Research Center and the NASA Center for Climate Simulation at Goddard Space Flight Center. We acknowledge the \psp\ mission for use of the data, which are publicly available at the \href{https://spdf.gsfc.nasa.gov/}{NASA Space Physics Data Facility}. This work utilizes data produced collaboratively between AFRL/ADAPT and NSO/NISP.
\end{acknowledgments}

\appendix

\section{Structural Similarity of 2nd-Order Correlation Functions Measured by PSP}\label{sec:app}

The \cite{usmanov2018} model employs a turbulence transport model which adapts and extends the \cite{breech2008turbulence} model. Rather than engage the full potential complexity of 
non-WKB transport theory
\citep{zhou1990transport,matthaeus1994JGRwkb}, these models adopt simplifying assumptions that reduce substantially the number and complexity of the full models. One key approximation is  
so-called {\it structural similarity}. 
In its most general form, this approximation 
asserts that
the second-order correlation functions
\citep{oughton1997PRE} involving magnetic field, velocity, and both Elsasser fields, are of same functional forms, apart from their respective energy normalizations, in a given sample of the turbulence. 
A corollary is that the correlation (or coherence) scales of these fields are identical. 
It is known that 
these correlation scales cannot be equated in general
\citep[see, e.g.,][]{wan2012JFM697,dosch2013SWproc}, and 
related transport models have been developed that retain an additional number of independent lengths scales \citep{Zank2017ApJ835,adhikari2020ApJ}. 
Nevertheless, it is instructive to examine whether 
in the present context the simpler
assumption we employ 
provides reasonable approximations to correlation lengths and correlation functions observed by PSP.  

\begin{figure}
    \centering
    \includegraphics[width=.32\textwidth]{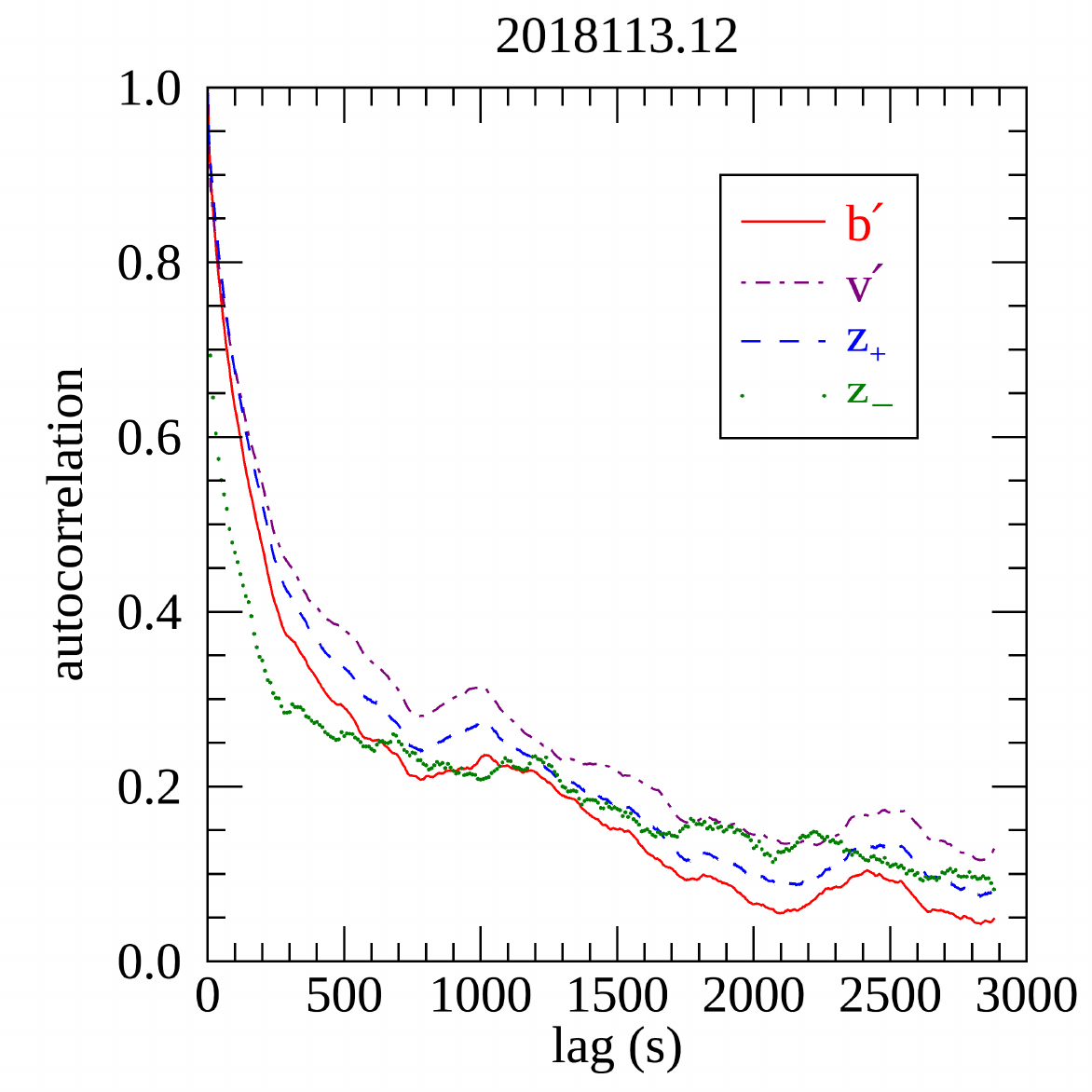}
    \includegraphics[width=.32\textwidth]{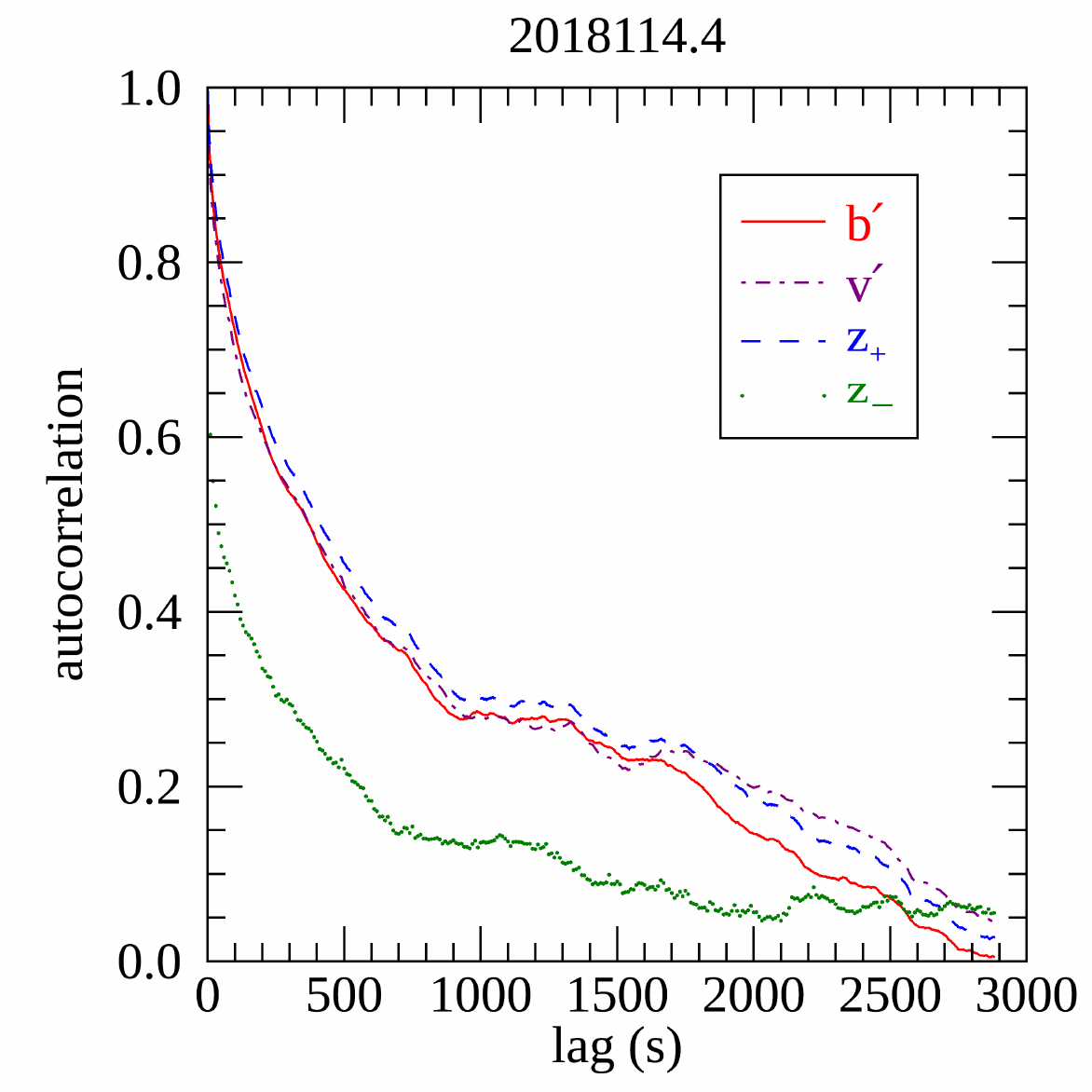}
    \includegraphics[width=.32\textwidth]{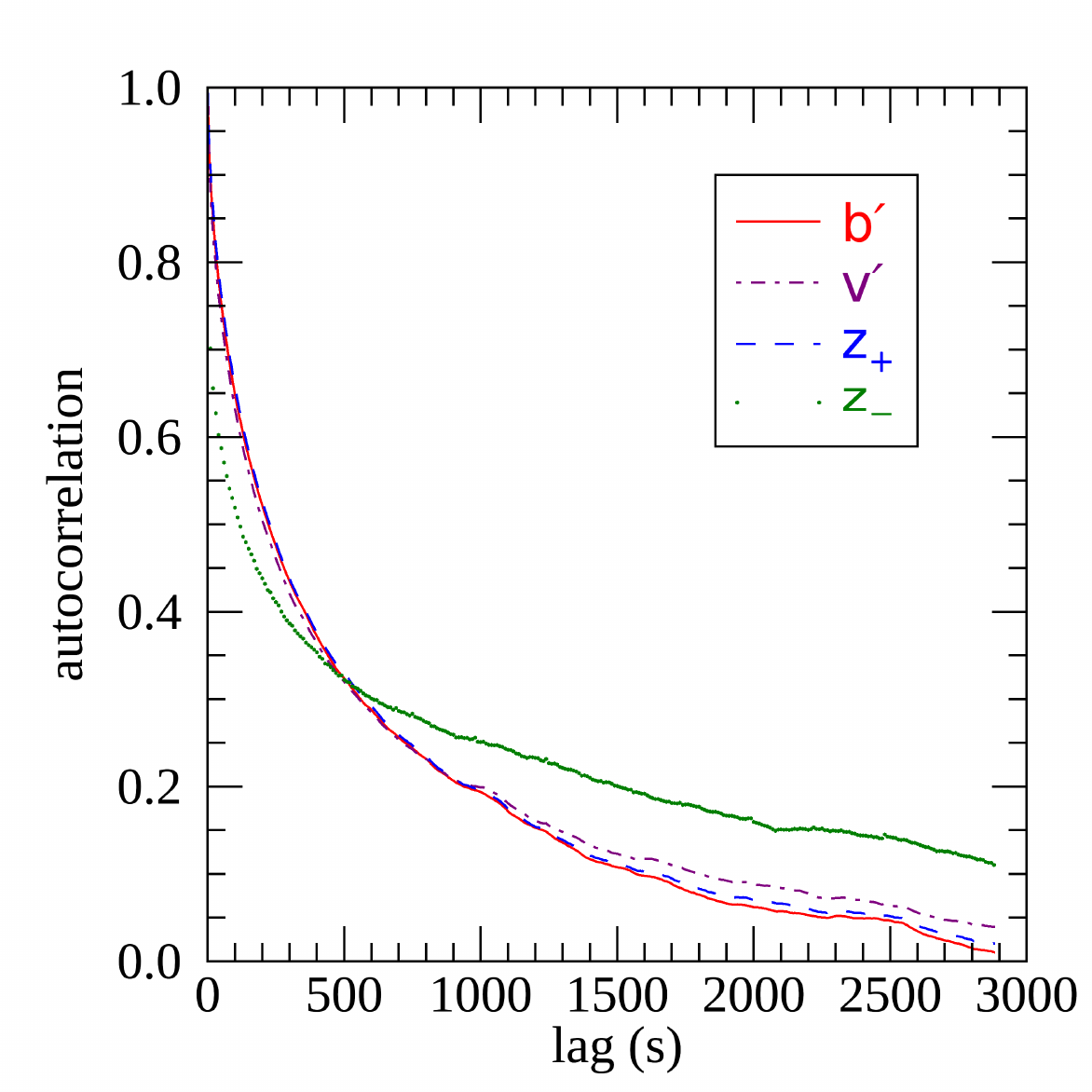}
    \caption{Autocorrelation functions (ACFs) of magnetic \(\bm{b}'\) and velocity \(\bm{v}'\) fluctuations, and Elsasser variables \(\bm{z}_\pm\). \textit{Left and Middle}: ACFs computed from 8-hour intervals centered at 12:00 UTC on 2018 November 3 and 04:00 UTC on 2018 November 4, respectively. Labels on top of panels show date and time in the format YYYYMMD.H, where Y is year, M is month, D is day, and H is hour. \textit{Right}: ACFs averaged over seventeen 8-hour intervals near PSP's first perihelion, covering the times 00:00 UTC on 2018 November 3 to 16:00 UTC on 2018 November 8. The average correlation times (computed by averaging the correlation times obtained from each interval) are 453, 433, 443, and 535 seconds, for \(\bm{b}',~\bm{v}',~\bm{z}_+,\) and \(\bm{z}_-\),  respectively.}
    \label{fig:app1}
\end{figure}

To this end we computed the four relevant autocorrelations \citep[see, e.g.,][]{matthaeus1982JGR},
 for magnetic \(\bm{b}'\) and velocity \(\bm{v}'\) fluctuations, and Elsasser variables \(\bm{z}_\pm\), 
for seventeen 8-hour intervals covering a few days around PSP's first perihelion (3rd Nov 2018 to 8th Nov 2018). Examples are shown in the left and middle panels of Figure \ref{fig:app1}, portrayed as 
functions of temporal lag. For the eight hour sample on 2018 Nov 3 (left panel), observe that all four correlations are of very similar form. In the interval shown from 2018 Nov 4 (middle panel), three of the four are very similar to one another, while the fourth, the minority Elsasser amplitude, displays a more rapidly decaying 
autocorrelation than the others. This behavior is representative of averages as well. The right panel shows the averages of the 
four autocorrelations over several days near first PSP 
perihelion. Again, three of the four are extremely similar to one another, the exception being the minority amplitude associated with 
``inward'' fluctuations, and in that case a modest departure 
from the others is observed. As one would expect, the 
computed average correlation times do not differ greatly, even if significant differences can be found in some subintervals. The average correlation times (computed by averaging the correlation times obtained from each interval) are 453, 433, 443, and 535 seconds, for \(\bm{b}',~\bm{v}',~\bm{z}_+,\) and \(\bm{z}_-\),  
respectively. This demonstrates that for at least this limited 
sample, it is unlikely that great inaccuracy 
in the evolution of energy is likely to be introduced 
by the assumption of a single similarity scale, although this
may have a moderate impact on the 
decay rate of the smaller minority Elsasser amplitude. We close with the caveat that the present analysis is preliminary; a more detailed investigation of such structural similarity is under way.
%



\begin{thebibliography}{}
\expandafter\ifx\csname natexlab\endcsname\relax\def\natexlab#1{#1}\fi
\providecommand{\url}[1]{\href{#1}{#1}}
\providecommand{\dodoi}[1]{doi:~\href{http://doi.org/#1}{\nolinkurl{#1}}}
\providecommand{\doeprint}[1]{\href{http://ascl.net/#1}{\nolinkurl{http://ascl.net/#1}}}
\providecommand{\doarXiv}[1]{\href{https://arxiv.org/abs/#1}{\nolinkurl{https://arxiv.org/abs/#1}}}

\bibitem[{{Adhikari} {et~al.}(2020){Adhikari}, {Zank}, \&
  {Zhao}}]{adhikari2020ApJ}
{Adhikari}, L., {Zank}, G.~P., \& {Zhao}, L.~L. 2020, \apj, 901, 102,
  \dodoi{10.3847/1538-4357/abb132}

\bibitem[{{Arge} {et~al.}(2010){Arge}, {Henney}, {Koller}, {Compeau}, {Young},
  {MacKenzie}, {Fay}, \& {Harvey}}]{arge2010AIPC}
{Arge}, C.~N., {Henney}, C.~J., {Koller}, J., {et~al.} 2010, in American
  Institute of Physics Conference Series, Vol. 1216, Twelfth International
  Solar Wind Conference, ed. M.~{Maksimovic}, K.~{Issautier},
  N.~{Meyer-Vernet}, M.~{Moncuquet}, \& F.~{Pantellini}, 343--346,
  \dodoi{10.1063/1.3395870}

\bibitem[{{Badman} {et~al.}(2020){Badman}, {Bale}, {Mart{\'\i}nez Oliveros},
  {Panasenco}, {Velli}, {Stansby}, {Buitrago-Casas}, {R{\'e}ville}, {Bonnell},
  {Case}, {Dudok de Wit}, {Goetz}, {Harvey}, {Kasper}, {Korreck}, {Larson},
  {Livi}, {MacDowall}, {Malaspina}, {Pulupa}, {Stevens}, \&
  {Whittlesey}}]{badman2020ApJS}
{Badman}, S.~T., {Bale}, S.~D., {Mart{\'\i}nez Oliveros}, J.~C., {et~al.} 2020,
  \apjs, 246, 23, \dodoi{10.3847/1538-4365/ab4da7}

\bibitem[{{Bale} {et~al.}(2016){Bale}, {Goetz}, {Harvey}, {Turin}, {Bonnell},
  {Dudok de Wit}, {Ergun}, {MacDowall}, {Pulupa}, {Andre}, {Bolton},
  {Bougeret}, {Bowen}, {Burgess}, {Cattell}, {Chandran}, {Chaston}, {Chen},
  {Choi}, {Connerney}, {Cranmer}, {Diaz-Aguado}, {Donakowski}, {Drake},
  {Farrell}, {Fergeau}, {Fermin}, {Fischer}, {Fox}, {Glaser}, {Goldstein},
  {Gordon}, {Hanson}, {Harris}, {Hayes}, {Hinze}, {Hollweg}, {Horbury},
  {Howard}, {Hoxie}, {Jannet}, {Karlsson}, {Kasper}, {Kellogg}, {Kien},
  {Klimchuk}, {Krasnoselskikh}, {Krucker}, {Lynch}, {Maksimovic}, {Malaspina},
  {Marker}, {Martin}, {Martinez-Oliveros}, {McCauley}, {McComas}, {McDonald},
  {Meyer-Vernet}, {Moncuquet}, {Monson}, {Mozer}, {Murphy}, {Odom},
  {Oliverson}, {Olson}, {Parker}, {Pankow}, {Phan}, {Quataert}, {Quinn},
  {Ruplin}, {Salem}, {Seitz}, {Sheppard}, {Siy}, {Stevens}, {Summers}, {Szabo},
  {Timofeeva}, {Vaivads}, {Velli}, {Yehle}, {Werthimer}, \&
  {Wygant}}]{bale2016SSR}
{Bale}, S.~D., {Goetz}, K., {Harvey}, P.~R., {et~al.} 2016, \ssr, 204, 49,
  \dodoi{10.1007/s11214-016-0244-5}

\bibitem[{{Bale} {et~al.}(2019){Bale}, {Badman}, {Bonnell}, {Bowen}, {Burgess},
  {Case}, {Cattell}, {Chandran}, {Chaston}, {Chen}, {Drake}, {de Wit},
  {Eastwood}, {Ergun}, {Farrell}, {Fong}, {Goetz}, {Goldstein}, {Goodrich},
  {Harvey}, {Horbury}, {Howes}, {Kasper}, {Kellogg}, {Klimchuk}, {Korreck},
  {Krasnoselskikh}, {Krucker}, {Laker}, {Larson}, {MacDowall}, {Maksimovic},
  {Malaspina}, {Martinez-Oliveros}, {McComas}, {Meyer-Vernet}, {Moncuquet},
  {Mozer}, {Phan}, {Pulupa}, {Raouafi}, {Salem}, {Stansby}, {Stevens}, {Szabo},
  {Velli}, {Woolley}, \& {Wygant}}]{bale2019Nat}
{Bale}, S.~D., {Badman}, S.~T., {Bonnell}, J.~W., {et~al.} 2019, \nat, 576,
  237, \dodoi{10.1038/s41586-019-1818-7}

\bibitem[{Bandyopadhyay {et~al.}(2018)Bandyopadhyay, Oughton, Wan, Matthaeus,
  Chhiber, \& Parashar}]{bandyopadhyay2018prx}
Bandyopadhyay, R., Oughton, S., Wan, M., {et~al.} 2018, Phys. Rev. X, 8,
  041052, \dodoi{10.1103/PhysRevX.8.041052}

\bibitem[{{Bandyopadhyay} {et~al.}(2018){Bandyopadhyay}, {Chasapis}, {Chhiber},
  {Parashar}, {Maruca}, {Matthaeus}, {Schwartz}, {Eriksson}, {Le Contel},
  {Breuillard}, {Burch}, {Moore}, {Pollock}, {Giles}, {Paterson}, {Dorelli},
  {Gershman}, {Torbert}, {Russell}, \& {Strangeway}}]{bandyopadhyay2018filter}
{Bandyopadhyay}, R., {Chasapis}, A., {Chhiber}, R., {et~al.} 2018, \apj, 866,
  81, \dodoi{10.3847/1538-4357/aade93}

\bibitem[{{Bandyopadhyay} {et~al.}(2020{\natexlab{a}}){Bandyopadhyay},
  {Goldstein}, {Maruca}, {Matthaeus}, {Parashar}, {Ruffolo}, {Chhiber},
  {Usmanov}, {Chasapis}, {Qudsi}, {Bale}, {Bonnell}, {Dudok de Wit}, {Goetz},
  {Harvey}, {MacDowall}, {Malaspina}, {Pulupa}, {Kasper}, {Korreck}, {Case},
  {Stevens}, {Whittlesey}, {Larson}, {Livi}, {Klein}, {Velli}, \&
  {Raouafi}}]{bandyopadhyay2020ApJS_cascade}
{Bandyopadhyay}, R., {Goldstein}, M.~L., {Maruca}, B.~A., {et~al.}
  2020{\natexlab{a}}, \apjs, 246, 48, \dodoi{10.3847/1538-4365/ab5dae}

\bibitem[{{Bandyopadhyay} {et~al.}(2020{\natexlab{b}}){Bandyopadhyay}, {Yang},
  {Matthaeus}, {Chasapis}, {Parashar}, {Russell}, {Strangeway}, {Torbert},
  {Giles}, {Gershman}, {Pollock}, {Moore}, \&
  {Burch}}]{bandyopadhyay2020ApJL_curvature}
{Bandyopadhyay}, R., {Yang}, Y., {Matthaeus}, W.~H., {et~al.}
  2020{\natexlab{b}}, \apjl, 893, L25, \dodoi{10.3847/2041-8213/ab846e}

\bibitem[{{Barnes}(1979)}]{barnes1979inbook}
{Barnes}, A. 1979, {Hydromagnetic waves and turbulence in the solar wind}, ed.
  E.~N. {Parker}, C.~F. {Kennel}, \& L.~J. {Lanzerotti} (National Academies
  Press), 249--319

\bibitem[{{Belcher} \& {Davis}(1971)}]{belcher1971JGR}
{Belcher}, J.~W., \& {Davis}, Jr., L. 1971, \jgr, 76, 3534,
  \dodoi{10.1029/JA076i016p03534}

\bibitem[{{Breech} {et~al.}(2008){Breech}, {Matthaeus}, {Minnie}, {Bieber},
  {Oughton}, {Smith}, \& {Isenberg}}]{breech2008turbulence}
{Breech}, B., {Matthaeus}, W.~H., {Minnie}, J., {et~al.} 2008, Journal of
  Geophysical Research (Space Physics), 113, A08105,
  \dodoi{10.1029/2007JA012711}

\bibitem[{{Case} {et~al.}(2020){Case}, {Kasper}, {Stevens}, {Korreck},
  {Paulson}, {Daigneau}, {Caldwell}, {Freeman}, {Henry}, {Klingensmith},
  {Bookbinder}, {Robinson}, {Berg}, {Tiu}, {Wright}, {Reinhart}, {Curtis},
  {Ludlam}, {Larson}, {Whittlesey}, {Livi}, {Klein}, \&
  {Martinovi{\'c}}}]{case2020ApJS}
{Case}, A.~W., {Kasper}, J.~C., {Stevens}, M.~L., {et~al.} 2020, \apjs, 246,
  43, \dodoi{10.3847/1538-4365/ab5a7b}

\bibitem[{{Chen} {et~al.}(2020){Chen}, {Bale}, {Bonnell}, {Borovikov}, {Bowen},
  {Burgess}, {Case}, {Chandran}, {de Wit}, {Goetz}, {Harvey}, {Kasper},
  {Klein}, {Korreck}, {Larson}, {Livi}, {MacDowall}, {Malaspina}, {Mallet},
  {McManus}, {Moncuquet}, {Pulupa}, {Stevens}, \& {Whittlesey}}]{chen2020ApJS}
{Chen}, C.~H.~K., {Bale}, S.~D., {Bonnell}, J.~W., {et~al.} 2020, \apjs, 246,
  53, \dodoi{10.3847/1538-4365/ab60a3}

\bibitem[{{Chhiber} {et~al.}(2021{\natexlab{a}}){Chhiber}, {Ruffolo},
  {Matthaeus}, {Usmanov}, {Tooprakai}, {Chuychai}, \&
  {Goldstein}}]{chhiber2021ApJ_flrw}
{Chhiber}, R., {Ruffolo}, D., {Matthaeus}, W.~H., {et~al.} 2021{\natexlab{a}},
  \apj, 908, 174, \dodoi{10.3847/1538-4357/abd7f0}

\bibitem[{{Chhiber} {et~al.}(2017){Chhiber}, {Subedi}, {Usmanov}, {Matthaeus},
  {Ruffolo}, {Goldstein}, \& {Parashar}}]{chhiber2017ApJS230}
{Chhiber}, R., {Subedi}, P., {Usmanov}, A.~V., {et~al.} 2017, \apjs, 230, 21,
  \dodoi{10.3847/1538-4365/aa74d2}

\bibitem[{{Chhiber} {et~al.}(2016){Chhiber}, {Usmanov}, {Matthaeus}, \&
  {Goldstein}}]{chhiber2016solar}
{Chhiber}, R., {Usmanov}, A., {Matthaeus}, W., \& {Goldstein}, M. 2016, \apj,
  821, 34, \dodoi{10.3847/0004-637X/821/1/34}

\bibitem[{{Chhiber} {et~al.}(2018){Chhiber}, {Usmanov}, {DeForest},
  {Matthaeus}, {Parashar}, \& {Goldstein}}]{chhiber2018apjl}
{Chhiber}, R., {Usmanov}, A.~V., {DeForest}, C.~E., {et~al.} 2018, \apjl, 856,
  L39, \dodoi{10.3847/2041-8213/aab843}

\bibitem[{{Chhiber} {et~al.}(2019{\natexlab{a}}){Chhiber}, {Usmanov},
  {Matthaeus}, \& {Goldstein}}]{chhiber2019psp1}
{Chhiber}, R., {Usmanov}, A.~V., {Matthaeus}, W.~H., \& {Goldstein}, M.~L.
  2019{\natexlab{a}}, \apjs, 241, 11, \dodoi{10.3847/1538-4365/ab0652}

\bibitem[{{Chhiber} {et~al.}(2019{\natexlab{b}}){Chhiber}, {Usmanov},
  {Matthaeus}, {Parashar}, \& {Goldstein}}]{chhiber2019psp2}
{Chhiber}, R., {Usmanov}, A.~V., {Matthaeus}, W.~H., {Parashar}, T.~N., \&
  {Goldstein}, M.~L. 2019{\natexlab{b}}, \apjs, 242, 12,
  \dodoi{10.3847/1538-4365/ab16d7}

\bibitem[{{Chhiber} {et~al.}(2020){Chhiber}, {Goldstein}, {Maruca}, {Chasapis},
  {Matthaeus}, {Ruffolo}, {Bandyopadhyay}, {Parashar}, {Qudsi}, {de Wit},
  {Bale}, {Bonnell}, {Goetz}, {Harvey}, {MacDowall}, {Malaspina}, {Pulupa},
  {Kasper}, {Korreck}, {Case}, {Stevens}, {Whittlesey}, {Larson}, {Livi},
  {Velli}, \& {Raouafi}}]{chhiber2020ApJS}
{Chhiber}, R., {Goldstein}, M.~L., {Maruca}, B.~A., {et~al.} 2020, \apjs, 246,
  31, \dodoi{10.3847/1538-4365/ab53d2}

\bibitem[{{Chhiber} {et~al.}(2021{\natexlab{b}}){Chhiber}, {Matthaeus},
  {Cohen}, {Ruffolo}, {Sonsrettee}, {Tooprakai}, {Seripienlert}, {Chuychai},
  {Usmanov}, {Goldstein}, {McComas}, {Leske}, {Christian}, {Mewaldt},
  {Labrador}, {Szalay}, {Joyce}, {Giacalone}, {Schwadron}, {Mitchell}, {Hill},
  {Wiedenbeck}, {McNutt}, \& {Desai}}]{chhiber2021AA}
{Chhiber}, R., {Matthaeus}, W.~H., {Cohen}, C.~M.~S., {et~al.}
  2021{\natexlab{b}}, \aap, \dodoi{https://doi.org/10.1051/0004-6361/202039816}

\bibitem[{{Cranmer} {et~al.}(2009){Cranmer}, {Matthaeus}, {Breech}, \&
  {Kasper}}]{cranmer2009ApJ}
{Cranmer}, S.~R., {Matthaeus}, W.~H., {Breech}, B.~A., \& {Kasper}, J.~C. 2009,
  \apj, 702, 1604, \dodoi{10.1088/0004-637X/702/2/1604}

\bibitem[{{Dasso} {et~al.}(2005){Dasso}, {Milano}, {Matthaeus}, \&
  {Smith}}]{dasso2005ApJ}
{Dasso}, S., {Milano}, L.~J., {Matthaeus}, W.~H., \& {Smith}, C.~W. 2005,
  \apjl, 635, L181, \dodoi{10.1086/499559}

\bibitem[{{de K\'arm\'an} \& {Howarth}(1938)}]{karman1938prsl}
{de K\'arm\'an}, T., \& {Howarth}, L. 1938, Proceedings of the Royal Society of
  London Series A, 164, 192, \dodoi{10.1098/rspa.1938.0013}

\bibitem[{{DeForest} {et~al.}(2019){DeForest}, {Gibson}, {Beasley},
  {Colaninno}, {Killough}, {Kosmann}, {Laurent}, \&
  {McMullin}}]{deforest2019AGU_PUNCH}
{DeForest}, C.~E., {Gibson}, S.~E., {Beasley}, M., {et~al.} 2019, in AGU Fall
  Meeting Abstracts, Vol. 2019, SH43B--06

\bibitem[{{Dosch} {et~al.}(2013){Dosch}, {Adhikari}, \&
  {Zank}}]{dosch2013SWproc}
{Dosch}, A., {Adhikari}, L., \& {Zank}, G.~P. 2013, in American Institute of
  Physics Conference Series, Vol. 1539, Solar Wind 13, ed. G.~P. {Zank},
  J.~{Borovsky}, R.~{Bruno}, J.~{Cirtain}, S.~{Cranmer}, H.~{Elliott},
  J.~{Giacalone}, W.~{Gonzalez}, G.~{Li}, E.~{Marsch}, E.~{Moebius},
  N.~{Pogorelov}, J.~{Spann}, \& O.~{Verkhoglyadova}, 155--158,
  \dodoi{10.1063/1.4811011}

\bibitem[{{Dudok de Wit} {et~al.}(2020){Dudok de Wit}, {Krasnoselskikh},
  {Bale}, {Bonnell}, {Bowen}, {Chen}, {Froment}, {Goetz}, {Harvey},
  {Jagarlamudi}, {Larosa}, {MacDowall}, {Malaspina}, {Matthaeus}, {Pulupa},
  {Velli}, \& {Whittlesey}}]{DudokDeWit2020ApJS}
{Dudok de Wit}, T., {Krasnoselskikh}, V.~V., {Bale}, S.~D., {et~al.} 2020,
  \apjs, 246, 39, \dodoi{10.3847/1538-4365/ab5853}

\bibitem[{{Els\"asser}(1950)}]{elsasser1950PhRv}
{Els\"asser}, W.~M. 1950, Physical Review, 79, 183,
  \dodoi{10.1103/PhysRev.79.183}

\bibitem[{{Fox} {et~al.}(2016){Fox}, {Velli}, {Bale}, {Decker}, {Driesman},
  {Howard}, {Kasper}, {Kinnison}, {Kusterer}, {Lario}, {Lockwood}, {McComas},
  {Raouafi}, \& {Szabo}}]{fox2016SSR}
{Fox}, N.~J., {Velli}, M.~C., {Bale}, S.~D., {et~al.} 2016, \ssr, 204, 7,
  \dodoi{10.1007/s11214-015-0211-6}

\bibitem[{{Fr{\"a}nz} \& {Harper}(2002)}]{franz2002pss}
{Fr{\"a}nz}, M., \& {Harper}, D. 2002, \planss, 50, 217,
  \dodoi{10.1016/S0032-0633(01)00119-2}

\bibitem[{{Guo} \& {Florinski}(2016)}]{guo2016ApJ}
{Guo}, X., \& {Florinski}, V. 2016, \apj, 826, 65,
  \dodoi{10.3847/0004-637X/826/1/65}

\bibitem[{{Hartle} \& {Sturrock}(1968)}]{hartle1968ApJ151}
{Hartle}, R.~E., \& {Sturrock}, P.~A. 1968, \apj, 151, 1155,
  \dodoi{10.1086/149513}

\bibitem[{{Hollweg}(1974)}]{hollweg1974JGR79}
{Hollweg}, J.~V. 1974, \jgr, 79, 3845, \dodoi{10.1029/JA079i025p03845}

\bibitem[{{Hollweg}(1976)}]{hollweg1976JGR}
---. 1976, \jgr, 81, 1649, \dodoi{10.1029/JA081i010p01649}

\bibitem[{{Hossain} {et~al.}(1995){Hossain}, {Gray}, {Pontius}, {Matthaeus}, \&
  {Oughton}}]{hossain1995PhFl}
{Hossain}, M., {Gray}, P.~C., {Pontius}, Jr., D.~H., {Matthaeus}, W.~H., \&
  {Oughton}, S. 1995, Physics of Fluids, 7, 2886, \dodoi{10.1063/1.868665}

\bibitem[{{Howard} {et~al.}(2019){Howard}, {Vourlidas}, {Bothmer}, {Colaninno},
  {DeForest}, {Gallagher}, {Hall}, {Hess}, {Higginson}, {Korendyke},
  {Kouloumvakos}, {Lamy}, {Liewer}, {Linker}, {Linton}, {Penteado}, {Plunkett},
  {Poirier}, {Raouafi}, {Rich}, {Rochus}, {Rouillard}, {Socker}, {Stenborg},
  {Thernisien}, \& {Viall}}]{howard2019Nat}
{Howard}, R.~A., {Vourlidas}, A., {Bothmer}, V., {et~al.} 2019, \nat, 576, 232,
  \dodoi{10.1038/s41586-019-1807-x}

\bibitem[{{Isaacs} {et~al.}(2015){Isaacs}, {Tessein}, \&
  {Matthaeus}}]{isaacs2015JGR120}
{Isaacs}, J.~J., {Tessein}, J.~A., \& {Matthaeus}, W.~H. 2015, Journal of
  Geophysical Research (Space Physics), 120, 868, \dodoi{10.1002/2014JA020661}

\bibitem[{{Isenberg}(1986)}]{isenberg1986JGR}
{Isenberg}, P.~A. 1986, \jgr, 91, 9965, \dodoi{10.1029/JA091iA09p09965}

\bibitem[{{Kasper} {et~al.}(2016){Kasper}, {Abiad}, {Austin}, {Balat-Pichelin},
  {Bale}, {Belcher}, {Berg}, {Bergner}, {Berthomier}, {Bookbinder}, {Brodu},
  {Caldwell}, {Case}, {Chandran}, {Cheimets}, {Cirtain}, {Cranmer}, {Curtis},
  {Daigneau}, {Dalton}, {Dasgupta}, {DeTomaso}, {Diaz-Aguado}, {Djordjevic},
  {Donaskowski}, {Effinger}, {Florinski}, {Fox}, {Freeman}, {Gallagher},
  {Gary}, {Gauron}, {Gates}, {Goldstein}, {Golub}, {Gordon}, {Gurnee}, {Guth},
  {Halekas}, {Hatch}, {Heerikuisen}, {Ho}, {Hu}, {Johnson}, {Jordan},
  {Korreck}, {Larson}, {Lazarus}, {Li}, {Livi}, {Ludlam}, {Maksimovic},
  {McFadden}, {Marchant}, {Maruca}, {McComas}, {Messina}, {Mercer}, {Park},
  {Peddie}, {Pogorelov}, {Reinhart}, {Richardson}, {Robinson}, {Rosen},
  {Skoug}, {Slagle}, {Steinberg}, {Stevens}, {Szabo}, {Taylor}, {Tiu}, {Turin},
  {Velli}, {Webb}, {Whittlesey}, {Wright}, {Wu}, \& {Zank}}]{kasper2016SSR}
{Kasper}, J.~C., {Abiad}, R., {Austin}, G., {et~al.} 2016, \ssr, 204, 131,
  \dodoi{10.1007/s11214-015-0206-3}

\bibitem[{{Kasper} {et~al.}(2019){Kasper}, {Bale}, {Belcher}, {Berthomier},
  {Case}, {Chandran}, {Curtis}, {Gallagher}, {Gary}, {Golub}, {Halekas}, {Ho},
  {Horbury}, {Hu}, {Huang}, {Klein}, {Korreck}, {Larson}, {Livi}, {Maruca},
  {Lavraud}, {Louarn}, {Maksimovic}, {Martinovic}, {McGinnis}, {Pogorelov},
  {Richardson}, {Skoug}, {Steinberg}, {Stevens}, {Szabo}, {Velli},
  {Whittlesey}, {Wright}, {Zank}, {MacDowall}, {McComas}, {McNutt}, {Pulupa},
  {Raouafi}, \& {Schwadron}}]{kasper2019Nat}
{Kasper}, J.~C., {Bale}, S.~D., {Belcher}, J.~W., {et~al.} 2019, \nat, 576,
  228, \dodoi{10.1038/s41586-019-1813-z}

\bibitem[{Matsumoto \& Nishimura(1998)}]{matsumoto1998ACM}
Matsumoto, M., \& Nishimura, T. 1998, ACM Trans. Model. Comput. Simul., 8,
  3–30, \dodoi{10.1145/272991.272995}

\bibitem[{{Matthaeus} \& {Goldstein}(1982)}]{matthaeus1982JGR}
{Matthaeus}, W.~H., \& {Goldstein}, M.~L. 1982, \jgr, 87, 6011,
  \dodoi{10.1029/JA087iA08p06011}

\bibitem[{{Matthaeus} \& {Goldstein}(1986)}]{matthaeus1986prl}
---. 1986, Physical Review Letters, 57, 495, \dodoi{10.1103/PhysRevLett.57.495}

\bibitem[{{Matthaeus} {et~al.}(1990){Matthaeus}, {Goldstein}, \&
  {Roberts}}]{matthaeus1990JGR}
{Matthaeus}, W.~H., {Goldstein}, M.~L., \& {Roberts}, D.~A. 1990, \jgr, 95,
  20673, \dodoi{10.1029/JA095iA12p20673}

\bibitem[{{Matthaeus} {et~al.}(2004){Matthaeus}, {Minnie}, {Breech}, {Parhi},
  {Bieber}, \& {Oughton}}]{matthaeus2004grl}
{Matthaeus}, W.~H., {Minnie}, J., {Breech}, B., {et~al.} 2004, \grl, 31,
  L12803, \dodoi{10.1029/2004GL019645}

\bibitem[{{Matthaeus} {et~al.}(1994{\natexlab{a}}){Matthaeus}, {Oughton},
  {Pontius}, \& {Zhou}}]{matthaeus1994JGR}
{Matthaeus}, W.~H., {Oughton}, S., {Pontius}, Jr., D.~H., \& {Zhou}, Y.
  1994{\natexlab{a}}, \jgr, 99, 19, \dodoi{10.1029/94JA01233}

\bibitem[{{Matthaeus} {et~al.}(1998){Matthaeus}, {Smith}, \&
  {Oughton}}]{matthaeus1998JGR}
{Matthaeus}, W.~H., {Smith}, C.~W., \& {Oughton}, S. 1998, \jgr, 103, 6495,
  \dodoi{10.1029/97JA03729}

\bibitem[{{Matthaeus} {et~al.}(1996){Matthaeus}, {Zank}, \&
  {Oughton}}]{matthaeus1996jpp}
{Matthaeus}, W.~H., {Zank}, G.~P., \& {Oughton}, S. 1996, Journal of Plasma
  Physics, 56, 659, \dodoi{10.1017/S0022377800019516}

\bibitem[{{Matthaeus} {et~al.}(1994{\natexlab{b}}){Matthaeus}, {Zhou}, {Zank},
  \& {Oughton}}]{matthaeus1994JGRwkb}
{Matthaeus}, W.~H., {Zhou}, Y., {Zank}, G.~P., \& {Oughton}, S.
  1994{\natexlab{b}}, \jgr, 99, 23, \dodoi{10.1029/94JA02326}

\bibitem[{{McComas} {et~al.}(2018){McComas}, {Christian}, {Schwadron}, {Fox},
  {Westlake}, {Allegrini}, {Baker}, {Biesecker}, {Bzowski}, {Clark}, {Cohen},
  {Cohen}, {Dayeh}, {Decker}, {de Nolfo}, {Desai}, {Ebert}, {Elliott}, {Fahr},
  {Frisch}, {Funsten}, {Fuselier}, {Galli}, {Galvin}, {Giacalone},
  {Gkioulidou}, {Guo}, {Horanyi}, {Isenberg}, {Janzen}, {Kistler}, {Korreck},
  {Kubiak}, {Kucharek}, {Larsen}, {Leske}, {Lugaz}, {Luhmann}, {Matthaeus},
  {Mitchell}, {Moebius}, {Ogasawara}, {Reisenfeld}, {Richardson}, {Russell},
  {Sok{\'o}{\l}}, {Spence}, {Skoug}, {Sternovsky}, {Swaczyna}, {Szalay},
  {Tokumaru}, {Wiedenbeck}, {Wurz}, {Zank}, \& {Zirnstein}}]{mccomas2018SSR}
{McComas}, D.~J., {Christian}, E.~R., {Schwadron}, N.~A., {et~al.} 2018, \ssr,
  214, 116, \dodoi{10.1007/s11214-018-0550-1}

\bibitem[{{McComas} {et~al.}(2019){McComas}, {Christian}, {Cohen}, {Cummings},
  {Davis}, {Desai}, {Giacalone}, {Hill}, {Joyce}, {Krimigis}, {Labrador},
  {Leske}, {Malandraki}, {Matthaeus}, {McNutt}, {Mewaldt}, {Mitchell},
  {Posner}, {Rankin}, {Roelof}, {Schwadron}, {Stone}, {Szalay}, {Wiedenbeck},
  {Bale}, {Kasper}, {Case}, {Korreck}, {MacDowall}, {Pulupa}, {Stevens}, \&
  {Rouillard}}]{mccomas2019Nat}
{McComas}, D.~J., {Christian}, E.~R., {Cohen}, C.~M.~S., {et~al.} 2019, \nat,
  576, 223, \dodoi{10.1038/s41586-019-1811-1}

\bibitem[{{McComb}(1990)}]{mccomb1990physics}
{McComb}, W.~D. 1990, {The Physics of Fluid Turbulence} (Clarendon Press
  Oxford)

\bibitem[{{McManus} {et~al.}(2020){McManus}, {Bowen}, {Mallet}, {Chen},
  {Chandran}, {Bale}, {Larson}, {Dudok de Wit}, {Kasper}, {Stevens},
  {Whittlesey}, {Livi}, {Korreck}, {Goetz}, {Harvey}, {Pulupa}, {MacDowall},
  {Malaspina}, {Case}, \& {Bonnell}}]{mcmanus2020ApJS}
{McManus}, M.~D., {Bowen}, T.~A., {Mallet}, A., {et~al.} 2020, \apjs, 246, 67,
  \dodoi{10.3847/1538-4365/ab6dce}

\bibitem[{{Miesch} {et~al.}(2015){Miesch}, {Matthaeus}, {Brandenburg},
  {Petrosyan}, {Pouquet}, {Cambon}, {Jenko}, {Uzdensky}, {Stone}, {Tobias},
  {Toomre}, \& {Velli}}]{miesch2015SSR194}
{Miesch}, M., {Matthaeus}, W., {Brandenburg}, A., {et~al.} 2015, \ssr, 194, 97,
  \dodoi{10.1007/s11214-015-0190-7}

\bibitem[{{Moradi} \& {Li}(2019)}]{moradi2019ApJ}
{Moradi}, A., \& {Li}, G. 2019, \apj, 887, 102,
  \dodoi{10.3847/1538-4357/ab4f68}

\bibitem[{{M{\"u}ller} {et~al.}(2013){M{\"u}ller}, {Marsden}, {St. Cyr}, \&
  {Gilbert}}]{muller2013SoPh}
{M{\"u}ller}, D., {Marsden}, R.~G., {St. Cyr}, O.~C., \& {Gilbert}, H.~R. 2013,
  \solphys, 285, 25, \dodoi{10.1007/s11207-012-0085-7}

\bibitem[{Oughton {et~al.}(2015)Oughton, Matthaeus, Wan, \&
  Osman}]{oughton2015philtran}
Oughton, S., Matthaeus, W., Wan, M., \& Osman, K. 2015, Phil. Trans. R. Soc. A,
  373, 20140152

\bibitem[{{Oughton} {et~al.}(1997){Oughton}, {R{\"a}dler}, \&
  {Matthaeus}}]{oughton1997PRE}
{Oughton}, S., {R{\"a}dler}, K.~H., \& {Matthaeus}, W.~H. 1997, \pre, 56, 2875,
  \dodoi{10.1103/PhysRevE.56.2875}

\bibitem[{{Padhye} {et~al.}(2001){Padhye}, {Smith}, \&
  {Matthaeus}}]{padhye2001JGR}
{Padhye}, N.~S., {Smith}, C.~W., \& {Matthaeus}, W.~H. 2001, \jgr, 106, 18635,
  \dodoi{10.1029/2000JA000293}

\bibitem[{{Parashar} {et~al.}(2020){Parashar}, {Goldstein}, {Maruca},
  {Matthaeus}, {Ruffolo}, {Bandyopadhyay}, {Chhiber}, {Chasapis}, {Qudsi},
  {Vech}, {Roberts}, {Bale}, {Bonnell}, {de Wit}, {Goetz}, {Harvey},
  {MacDowall}, {Malaspina}, {Pulupa}, {Kasper}, {Korreck}, {Case}, {Stevens},
  {Whittlesey}, {Larson}, {Livi}, {Velli}, \& {Raouafi}}]{parashar2020ApJS}
{Parashar}, T.~N., {Goldstein}, M.~L., {Maruca}, B.~A., {et~al.} 2020, \apjs,
  246, 58, \dodoi{10.3847/1538-4365/ab64e6}

\bibitem[{Pearson(2002)}]{pearson2002hampel}
Pearson, R.~K. 2002, IEEE Transactions on Control Systems Technology, 10, 55,
  \dodoi{10.1109/87.974338}

\bibitem[{{Perez} {et~al.}(2021){Perez}, {Bourouaine}, {Chen}, \&
  {Raouafi}}]{perez2021AA}
{Perez}, J.~C., {Bourouaine}, S., {Chen}, C. H.~K., \& {Raouafi}, N.~E. 2021,
  \aap, 650, A22, \dodoi{10.1051/0004-6361/202039879}

\bibitem[{{R{\'e}ville} {et~al.}(2020){R{\'e}ville}, {Velli}, {Panasenco},
  {Tenerani}, {Shi}, {Badman}, {Bale}, {Kasper}, {Stevens}, {Korreck},
  {Bonnell}, {Case}, {de Wit}, {Goetz}, {Harvey}, {Larson}, {Livi},
  {Malaspina}, {MacDowall}, {Pulupa}, \& {Whittlesey}}]{reville2020ApJS}
{R{\'e}ville}, V., {Velli}, M., {Panasenco}, O., {et~al.} 2020, \apjs, 246, 24,
  \dodoi{10.3847/1538-4365/ab4fef}

\bibitem[{{Riley} {et~al.}(2014){Riley}, {Ben-Nun}, {Linker}, {Mikic},
  {Svalgaard}, {Harvey}, {Bertello}, {Hoeksema}, {Liu}, \&
  {Ulrich}}]{riley2014SoPh}
{Riley}, P., {Ben-Nun}, M., {Linker}, J.~A., {et~al.} 2014, \solphys, 289, 769,
  \dodoi{10.1007/s11207-013-0353-1}

\bibitem[{{Roberts} {et~al.}(1987){Roberts}, {Goldstein}, {Klein}, \&
  {Matthaeus}}]{roberts1987JGRb}
{Roberts}, D.~A., {Goldstein}, M.~L., {Klein}, L.~W., \& {Matthaeus}, W.~H.
  1987, \jgr, 92, 12023, \dodoi{10.1029/JA092iA11p12023}

\bibitem[{{Roberts} {et~al.}(1992){Roberts}, {Goldstein}, {Matthaeus}, \&
  {Ghosh}}]{roberts1992jgr}
{Roberts}, D.~A., {Goldstein}, M.~L., {Matthaeus}, W.~H., \& {Ghosh}, S. 1992,
  \jgr, 97, 17, \dodoi{10.1029/92JA01144}

\bibitem[{{Ruffolo} {et~al.}(2020){Ruffolo}, {Matthaeus}, {Chhiber}, {Usmanov},
  {Yang}, {Bandyopadhyay}, {Parashar}, {Goldstein}, {DeForest}, {Wan},
  {Chasapis}, {Maruca}, {Velli}, \& {Kasper}}]{ruffolo2020ApJ}
{Ruffolo}, D., {Matthaeus}, W.~H., {Chhiber}, R., {et~al.} 2020, \apj, 902, 94,
  \dodoi{10.3847/1538-4357/abb594}

\bibitem[{{Ruiz} {et~al.}(2011){Ruiz}, {Dasso}, {Matthaeus}, {Marsch}, \&
  {Weygand}}]{ruiz2011JGR}
{Ruiz}, M.~E., {Dasso}, S., {Matthaeus}, W.~H., {Marsch}, E., \& {Weygand},
  J.~M. 2011, Journal of Geophysical Research (Space Physics), 116, A10102,
  \dodoi{10.1029/2011JA016697}

\bibitem[{{Ruiz} {et~al.}(2014){Ruiz}, {Dasso}, {Matthaeus}, \&
  {Weygand}}]{Ruiz2014SoPh}
{Ruiz}, M.~E., {Dasso}, S., {Matthaeus}, W.~H., \& {Weygand}, J.~M. 2014, Solar
  Physics, 289, 3917, \dodoi{10.1007/s11207-014-0531-9}

\bibitem[{{Spence}(2019)}]{spence2019AGU_helioswarm}
{Spence}, H.~E. 2019, in AGU Fall Meeting Abstracts, Vol. 2019, SH11B--04

\bibitem[{{Spitzer} \& {H{\"a}rm}(1953)}]{spitzer1953PhRv}
{Spitzer}, L., \& {H{\"a}rm}, R. 1953, Physical Review, 89, 977,
  \dodoi{10.1103/PhysRev.89.977}

\bibitem[{{Szabo} {et~al.}(2020){Szabo}, {Larson}, {Whittlesey}, {Stevens},
  {Lavraud}, {Phan}, {Wallace}, {Jones-Mecholsky}, {Arge}, {Badman},
  {Odstrcil}, {Pogorelov}, {Kim}, {Riley}, {Henney}, {Bale}, {Bonnell}, {Case},
  {Dudok de Wit}, {Goetz}, {Harvey}, {Kasper}, {Korreck}, {Koval}, {Livi},
  {MacDowall}, {Malaspina}, \& {Pulupa}}]{szabo2020ApJS}
{Szabo}, A., {Larson}, D., {Whittlesey}, P., {et~al.} 2020, \apjs, 246, 47,
  \dodoi{10.3847/1538-4365/ab5dac}

\bibitem[{{Thom}(1958)}]{thom1958MWR}
{Thom}, H.~C.~S. 1958, Monthly Weather Review, 86, 117

\bibitem[{{Usmanov} {et~al.}(2012){Usmanov}, {Goldstein}, \&
  {Matthaeus}}]{usmanov2012three}
{Usmanov}, A.~V., {Goldstein}, M.~L., \& {Matthaeus}, W.~H. 2012, \apj, 754,
  40, \dodoi{10.1088/0004-637X/754/1/40}

\bibitem[{{Usmanov} {et~al.}(2014){Usmanov}, {Goldstein}, \&
  {Matthaeus}}]{usmanov2014three}
---. 2014, \apj, 788, 43, \dodoi{10.1088/0004-637X/788/1/43}

\bibitem[{{Usmanov} {et~al.}(2016){Usmanov}, {Goldstein}, \&
  {Matthaeus}}]{usmanov2016four}
---. 2016, \apj, 820, 17, \dodoi{10.3847/0004-637X/820/1/17}

\bibitem[{{Usmanov} {et~al.}(2011){Usmanov}, {Matthaeus}, {Breech}, \&
  {Goldstein}}]{usmanov2011solar}
{Usmanov}, A.~V., {Matthaeus}, W.~H., {Breech}, B.~A., \& {Goldstein}, M.~L.
  2011, \apj, 727, 84, \dodoi{10.1088/0004-637X/727/2/84}

\bibitem[{{Usmanov} {et~al.}(2018){Usmanov}, {Matthaeus}, {Goldstein}, \&
  {Chhiber}}]{usmanov2018}
{Usmanov}, A.~V., {Matthaeus}, W.~H., {Goldstein}, M.~L., \& {Chhiber}, R.
  2018, \apj, 865, 25, \dodoi{10.3847/1538-4357/aad687}

\bibitem[{{Verscharen} {et~al.}(2019){Verscharen}, {Klein}, \&
  {Maruca}}]{Verscharen2019LRSP}
{Verscharen}, D., {Klein}, K.~G., \& {Maruca}, B.~A. 2019, Living Reviews in
  Solar Physics, 16, 5, \dodoi{10.1007/s41116-019-0021-0}

\bibitem[{{Wan} {et~al.}(2012){Wan}, {Oughton}, {Servidio}, \&
  {Matthaeus}}]{wan2012JFM697}
{Wan}, M., {Oughton}, S., {Servidio}, S., \& {Matthaeus}, W.~H. 2012, Journal
  of Fluid Mechanics, 697, 296, \dodoi{10.1017/jfm.2012.61}

\bibitem[{{Wiengarten} {et~al.}(2016){Wiengarten}, {Oughton}, {Engelbrecht},
  {Fichtner}, {Kleimann}, \& {Scherer}}]{wiengarten2016ApJ833}
{Wiengarten}, T., {Oughton}, S., {Engelbrecht}, N.~E., {et~al.} 2016, \apj,
  833, 17, \dodoi{10.3847/0004-637X/833/1/17}

\bibitem[{{Wu} {et~al.}(2013){Wu}, {Wan}, {Matthaeus}, {Shay}, \&
  {Swisdak}}]{wu2013prl}
{Wu}, P., {Wan}, M., {Matthaeus}, W.~H., {Shay}, M.~A., \& {Swisdak}, M. 2013,
  Physical Review Letters, 111, 121105, \dodoi{10.1103/PhysRevLett.111.121105}

\bibitem[{{Zank} {et~al.}(2017){Zank}, {Adhikari}, {Hunana}, {Shiota}, {Bruno},
  \& {Telloni}}]{Zank2017ApJ835}
{Zank}, G.~P., {Adhikari}, L., {Hunana}, P., {et~al.} 2017, \apj, 835, 147,
  \dodoi{10.3847/1538-4357/835/2/147}

\bibitem[{{Zank} {et~al.}(1996){Zank}, {Matthaeus}, \&
  {Smith}}]{zank1996evolution}
{Zank}, G.~P., {Matthaeus}, W.~H., \& {Smith}, C.~W. 1996, \jgr, 101, 17093,
  \dodoi{10.1029/96JA01275}

\bibitem[{{Zhou} \& {Matthaeus}(1990)}]{zhou1990transport}
{Zhou}, Y., \& {Matthaeus}, W.~H. 1990, \jgr, 95, 10291,
  \dodoi{10.1029/JA095iA07p10291}

\end{thebibliography}

\listofchanges
\end{document}